\documentclass[aps,prl,reprint, amsmath, amssymb,superscriptaddress]{revtex4-1}
\usepackage{bm}
\usepackage[retainorgcmds]{IEEEtrantools}
\usepackage{graphicx}
\usepackage{mathrsfs}
\usepackage{amsmath}
\usepackage{amssymb}
\usepackage{color}
\usepackage{amsfonts,txfonts}
\usepackage{nicefrac}
\usepackage{hyperref}

\DeclareMathOperator{\tr}{tr}

\newlength{\eqboxstorage}

\begin{document}

\title{Thermodynamics of weakly coherent collisional models}
\date{\today}
\author{Franklin L. S. Rodrigues}
\affiliation{Instituto de F\'isica da Universidade de S\~ao Paulo,  05314-970 S\~ao Paulo, Brazil}
\author{Gabriele De Chiara}
\affiliation{Centre for Theoretical Atomic, Molecular and Optical Physics,
School of Mathematics and Physics, Queen's University Belfast, Belfast BT7 1NN, United Kingdom}
\author{Mauro Paternostro}
\affiliation{Centre for Theoretical Atomic, Molecular and Optical Physics,
School of Mathematics and Physics, Queen's University Belfast, Belfast BT7 1NN, United Kingdom}
\author{Gabriel T. Landi}
\email{gtlandi@if.usp.br}
\affiliation{Instituto de F\'isica da Universidade de S\~ao Paulo,  05314-970 S\~ao Paulo, Brazil}

\begin{abstract}

We introduce the idea of weakly coherent collisional models, where the elements of an environment interacting with a system of interest are prepared in states that are approximately thermal, but have an amount of coherence proportional to a short system-environment interaction time in a scenario akin to well-known collisional models. We show that, in the continuous-time limit, 
the model allows for a clear formulation of the first and second laws of thermodynamics, which are modified to include a non-trivial contribution related to quantum coherence.
Remarkably, we derive a bound showing that the degree of such coherence in the state of the elements of the environment represents a resource, which can be consumed to convert heat into an ordered (unitary-like) energy term  in the system, even though no work is performed in the global dynamics.
Our results therefore represent an instance where thermodynamics can be extended beyond thermal systems, opening the way for combining classical and quantum resources.  

\end{abstract}
\maketitle{}

%
%
%
%
The laws of thermodynamics provide operationally meaningful prescriptions on the tasks one may perform, given a set of available resources. 
The second law, in particular, sets strict bounds on the amount of work that can be extracted in a certain protocol.
Most  processes in Nature, however, are not thermodynamic and therefore do not enjoy such a simple and far reaching set of rules.
One is then  led to ask whether there exists scenarios \emph{``beyond thermal''} for which a clear set of thermodynamic rules can nonetheless be constructed.
This issue has recently been addressed, e.g., in the context of {\color{black}non-thermal} heat engines \cite{Scully2007,Dillenschneider2009a,Gardas2015},  squeezed thermal baths \cite{Manzano2016,Manzano2018b,Ronagel2014},  coherence amplification  \cite{Manzano2019}, information flows \cite{Ptaszynski2019a} and quantum resource theories \cite{Holmes2018a,Korzekwa2016,Baumer2018}. 
The question also acquires additional meaning in light of recent experimental demonstrations  that quantum effects can indeed be used as thermodynamic resources \cite{Micadei2017,Klaers2017a}.  

A framework that is particularly suited for addressing the thermodynamics of engineered reservoirs is that of collisional models (also called repeated interactions) \cite{Scarani2002,Ziman2002,Karevski2009,Giovannetti2012,Landi2014b,Strasberg2016,Barra2015,Pereira2018,Lorenzo2015,Lorenzo2015b,Pezzutto2016,Pezzutto2019,Cusumano2018,Englert2002}. 
They draw inspiration from Boltzmann's original \emph{Stosszahlansatz}: at any given interval of time, the system $S$ will only interact with a tiny fraction of the environment. 
For instance, in Brownian motion, a particle  interacts with only a few water molecules at a time.
Moreover, this interaction lasts for an extremely short time, after which the molecule moves on, never to return  \cite{Doob1954}. 
Since the environment is large, the next molecule to arrive will be completely uncorrelated from the previous one, so the  process repeats anew. 

\begin{figure}[t!]
\includegraphics[width=.4\textwidth]{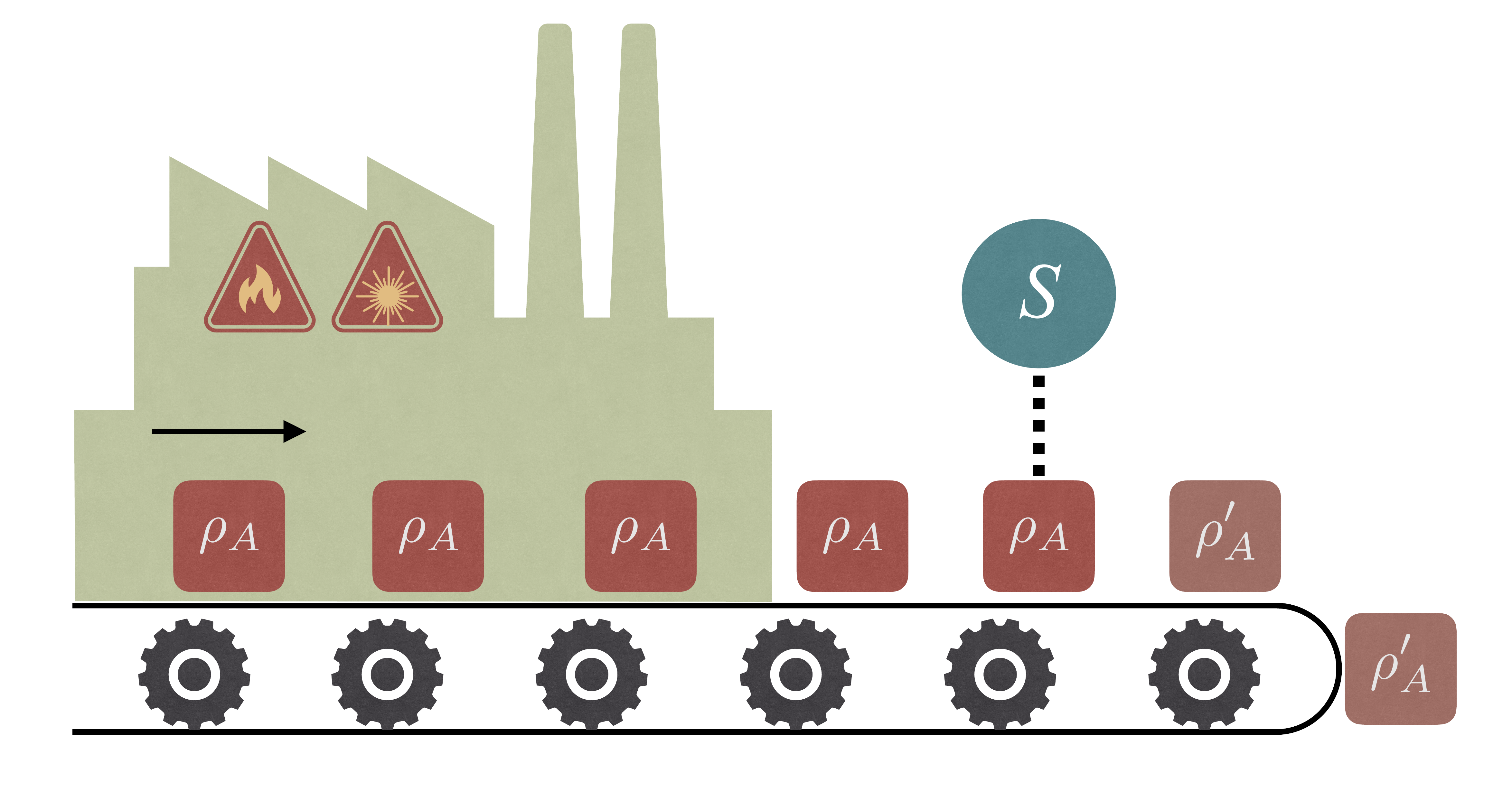}
\caption{\label{fig:drawing}
Basic setup of weakly coherent collisional models. 
The system is allowed to interact sequentially with a series of independent ancillae prepared in states $\rho_A$ which are close to being thermal, but have a small amount of coherence [cf. Eq.~(\ref{ancilla_state})]. 
}
\end{figure}

In the context of quantum systems, this can be viewed as the process depicted in Fig.~\ref{fig:drawing}, where the system $S$ interacts sequentially with a multi-party environment whose elements, henceforth dubbed {\it ancillae} $A_n$, are assumed to be mutually independent and prepared, in general, in arbitrary states. This generates a stroboscopic evolution for the reduced density matrix of the system, akin to a discrete-time Markov chain. 
A continuous-time description in terms of a Lindblad master equation can be derived in the short-time limit, provided  some assumptions are made about the system-ancilla interaction \cite{Englert2002,Karevski2009,Landi2014b}. 

When the ancillae are prepared in thermal states, it is possible to address quantitatively quantities of key thermodynamic relevance, from work to heat currents and entropy~\cite{Lorenzo2015,Lorenzo2015b,Pezzutto2016,Pezzutto2019,Cusumano2018}. 
This includes both the stroboscopic case, where formal relations can be drawn with the resource theory of athermality~\cite{Brandao2013,Brandao2015}, and the continuous-time limit~\cite{Strasberg2016,Santos2017}. 
The framework is also readily extended to systems coupled to multiple baths in an entirely consistent way \cite{Barra2015,DeChiara2018,Pereira2018}.
Conversely, when the state of the ancillae is not thermal, much less can be said about its thermodynamic properties.

An important contribution in this direction was given in Ref.~\cite{Strasberg2016,Manzano2017a}, which put forth a general framework  for describing the thermodynamics of collisional models. 
However, for general ancillary states, the second law of thermodynamics is expressed in terms of system-ancilla correlations and the changes in the state of the ancillae [cf. Eq.~(\ref{second_law}) below]. These quantities are rarely accessible in practice, which greatly limits the operational use of such formulations.

Motivated by this search for {``thermodynamics beyond thermal states"}, in this paper we draw a theoretical formulation of the  laws of thermodynamics for the class of \emph{weakly coherent collisional models} (Fig.~\ref{fig:drawing}), i.e.
situations where the ancillae are prepared in states that, albeit close to thermal ones, retain a small amount of coherence. This is a realistic case, as perfect thermal equilibrium is unlikely to be achieved in practice.
 
We show that, despite their weakness, the implications of such residual coherence for both the first and second law of thermodynamics are striking, in that non-trivial contributions to the continuous-time open dynamics arise to affect the phenomenology of energy exchanges between system and environment~\cite{Lorenzo2015}. In order to illustrate these features in a clear manner, we choose a scenario where no work is externally performed on the global system-ancilla compound~\cite{Barra2015,DeChiara2018,Pereira2018}, so that all changes in the energy of the system can be faithfully attributed to heat flowing from or into the environment. Despite this, we derive a bound showing how coherence in the ancillae (quantified by the relative entropy of coherence) is consumed to convert part of the heat into a coherent (work-like) term in the system. 

Our analysis thus entails that quantum coherence can embody a faithful resource in the energetics of open quantum systems~\cite{Santos2017}. Such resource can be consumed to transform disordered energy (heat) into ordered one (work), thus catalyzing the interconversion of thermodynamic energy exchanges of profoundly different nature, and paving the way to the control and steering of the thermodynamics of quantum processes.

%
%
%
%
{\bf \emph{Collisional models - }}
We begin by describing the general structure of collisional models.
A system $S$ interacts with an arbitrary number of environmental ancillae $A_1$, $A_2, \ldots$, all identically prepared in a certain state $\rho_A$. 
Each system-ancilla interaction lasts for a time $\tau$ and is governed by a unitary $U_{SA_n}$. 
The state of $S$ after its interaction with $A_n$ is embodied by the stroboscopic map 
\begin{equation}\label{stroboscopic_map}
\rho_S((n+1)\tau) =\tr_{A_n}\left(\rho'_{SA_n}\right)\equiv \tr_{A_n} \left[ U_{SA_n} \left(\rho_{S}(n\tau) \otimes \rho_{A} \right) U_{SA_n}^\dagger \right],
\end{equation}
where $\rho_{S}(n\tau)$ is the state of $S$ before the $n^{\rm th}$ system-ancilla interaction. 

Next, let $H_S$ and $H_{A_n}$ denote the free Hamiltonians of the system and ancillae. 
We define the heat exchanged in each interaction as the change in energy in the state of the ancilla~\cite{Reeb2014,Talkner2009,Goold2014b} $Q_{A_n}= \tr\big\{ H_{A_n} (\rho_{A_n}' - \rho_{A_n})\big\}$, where $\rho_{A_n}' = \tr_S \rho_{SA_n}'$.
Work is then defined as  the mismatch between $Q_{A_n}$ and the change in energy of the system, $\Delta E_n = \tr\big\{ H_S \big[\rho_S((n+1)\tau) - \rho_S(n\tau)\big]\big\}$, leading to the usual  first law of thermodynamics
\begin{equation}\label{first_law}
\Delta E_n = W_n  -Q_{A_n}.
\end{equation}
As the global dynamics is unitary, the definition of work in this case is unambiguous, being associated with the cost of switching the $S$-$A_n$ interaction on and off \cite{Barra2015,DeChiara2018,Pereira2018}. 
This work cost will be strictly zero whenever the system satisfies the condition~\cite{Brandao2013,Brandao2015}
$[U_{SA_n}, H_S + H_{A_n}]  = 0$,
which states strict energy conservation. 
In this case, Eq.~(\ref{first_law}) reduces to 
$\Delta E_n = - Q_{A_n}$,
which implies that all energy changes in the system can be unambiguously attributed to energy flowing to or from the ancillae. 
In order to highlight the role of quantum coherence, we shall assume this is the case throughout the paper.
The extension to the case where work is also present is straightforward.


The second law of thermodynamics for the map in Eq.~(\ref{stroboscopic_map}) can be expressed as the positivity of the entropy production in each stroke, defined as \cite{Strasberg2016,Manzano2017a}
\begin{equation}
 \label{second_law}
 \Sigma_n = \mathcal{I}(\rho_{SA_n}') + S(\rho^{\prime}_{A_n}||\rho_{A_n}),
\end{equation}
where $\mathcal{I}(\rho_{SA_n}') = S(\rho_S') + S(\rho_{A_n}') - S(\rho_{SA_n}')$ is the mutual information between $S$ and $A_n$ after their joint evolution, $S(\rho_{A_n}'||\rho_{A_n}) = \tr\big(\rho_{A_n}' \ln \rho_{A_n}' - \rho_{A_n}' \ln \rho_{A_n}\big)$ is the relative entropy between the initial and final states of $A_n$, and $S(\rho) = - \tr\left(\rho\ln \rho\right)$ is the von Neumann entropy. Eq.~(\ref{second_law}) quantifies the degree of irreversibility associated with tracing out the ancillae. 
It accounts  not only  for the system-ancilla correlations that are irretrievably lost in this process, but also for the change in state of the ancilla, represented by the last term in Eq.~(\ref{second_law}).
The two terms were recently compared in Refs.~\cite{Ptaszynski2019b,Pezzutto2016}, and in the context of Landauer's principle~\cite{Reeb2014}.

%
%
%
%
{\bf \emph{Continuous-time limit - }}
In the limit of small $\tau$, Eq.~(\ref{stroboscopic_map}) can be approximated by a Lindblad master equation. 
Such a limit 
requires a value of $\tau$ sufficiently small to allow 
us to approximate $\rho_S((n+1)\tau)-\rho_S(n\tau)$ as a sufficiently smooth derivative. 
Mathematically, in order to implement this, it is convenient to rescale the interaction potential $V_{SA_n}$ between $S$ and $A_n$ by a factor  $1/\sqrt{\tau}$~\cite{Englert2002,Karevski2009,Landi2014b}.
That is, one assumes that the total $S$-$A_n$ Hamiltonian is of the form
\begin{equation}\label{total_Hamiltonian}
H_{SA_n} = H_S + H_{A_n} + V_{SA_n}/\sqrt{\tau}
\end{equation}
with the unitary evolution $U_{SA_n} = \exp[- i \tau H_{SA_n}]$.
This kind of rescaling, which enables the performance of the continuous-time limit, is frequent in stochastic processes; e.g.~ in classical Brownian motion \footnote{
For instance,  the white noise appearing in the Langevin equation of Brownian motion acts for an infinitesimal time so, in order to be non-trivial, it has to also be infinitely strong  \cite{Coffey2004}.} or in the interaction with the radiation field~\cite{Ciccarello2017}.

%
%
%
%
{\bf \emph{Weakly coherent ancillae - }}
Finally, we specify the  state of the ancillae, which is the main feature of our construction. 
We assume that the ancillae are prepared in a state of the form 
\begin{equation}\label{ancilla_state}
\rho_{A} = \rho_A^\text{th} + \sqrt{\tau} \; \lambda \chi_A, 
\end{equation}
where $\rho_A^\text{th} = {e^{-\beta H_{A}}}/{Z_A}$ is a thermal state at the inverse temperature $\beta$ ($Z_A$ is the corresponding partition function).
Here $\chi_A$ is a Hermitian operator having no diagonal elements in the energy basis of $H_{A}$. 
Moreover, $\lambda$ is a control parameter that measures the magnitude of the coherences. 
Notice that the term ``weak coherences'' is used here in the sense that we are interested specifically in the case where $\tau \to 0$, in which case  the second term in~(\ref{ancilla_state}) is much smaller in magnitude than the first. 
For finite $\tau$, not all choices of $\chi_A$  lead to a positive semidefinite $\rho_A$.
However, in the limit $\tau\to 0$, these constraints are relaxed and  any form of $\chi_A$ having no diagonal entries becomes allowed. 

The scaling in Eq.~(\ref{ancilla_state}) 
highlights an interesting feature of coherent collisional models, namely that for a short $\tau$ and strong $V_{SA_n}$, even weak coherences already produce non-negligible contributions.

We use the unitary $U_{SA_n}$ generated by Eq.~(\ref{total_Hamiltonian}) and the state in Eq.~(\ref{ancilla_state}) in the map stated in Eq.~(\ref{stroboscopic_map}). We then expand the latter in power series of $\tau$ and take the limit $\tau \to 0$. 
This then leads to the quantum master equation (cf.~\cite{SupMat} for details)
\begin{equation}\label{M}
\dot{\rho}_S = -i [H_S + \lambda\; G, \rho_S] + D(\rho_S), 
\end{equation}
where $\dot{\rho}_S = \lim_{\tau \to 0} \big[\rho_S((n+1)\tau) - \rho_S(n\tau)\big]/\tau$. 
We also define 
\begin{equation}\label{D}
D(\rho_S) = -\tr_A [V_{SA}, [V_{SA}, \rho_S\otimes \rho_A^\text{th}]]/2, 
\end{equation}
representing the usual Lindblad dissipator associated with the thermal part  $\rho_A^\text{th}$, and 
\begin{equation}\label{G}
G = \tr_A (V_{SA} \chi_A)
\end{equation}
representing a new unitary contribution stemming from the coherent part of $\rho_A$.
In deriving Eq.~(\ref{M}) we have  assumed that $\tr_A (V_{SA} \rho_A^\text{th}) = 0$, as customary \cite{Rivas2012}. 
Eqs.~(\ref{M})-(\ref{G}) 
provide a general recipe for deriving quantum master equations in the presences of weak coherences. 
All one requires is the form of the system-ancilla interaction potential and the state of the ancillae. 
In the limit $\lambda \to 0$ one recovers the standard thermal master equation \cite{Englert2002,Karevski2009,Landi2014b,DeChiara2018}.


%
%
%
%
{\bf \emph{Eigenoperator interaction - }}
The physics of Eqs.~(\ref{M})-(\ref{G}) becomes clearer if one assumes a specific form for the interaction $V_{SA}$. 
A structure which is particularly illuminating, in light of the strict energy-conservation condition, is 
\begin{equation}\label{V_eigenops}
V_{SA} = \sum\limits_k g_k  L_k^\dagger A_k + h.c., 
\end{equation}
where $g_k$ are complex coefficients and $L_k$ and $A_k$ are eigenoperators for the system and ancilla respectively \cite{Breuer2007}. 
That is, they satisfy the conditions $[H_S, L_k] = - \omega_k L_k$ and $[H_A, A_k] = - \omega_k A_k$, for the same set of Bohr frequencies $\{\omega_k\}$. 
This means that they function as lowering and raising operators for the energy basis of $S$ and $A$. 
As both have the same $\omega_k$, all of the energy leaving the system enters an ancilla and viceversa, so that strict energy conservation is always satisfied.

The form taken by the dissipator in Eq.~(\ref{D}) when $V_{SA}$ is as given above is the standard thermal one 
\begin{equation}\label{D_eigenop}
D(\rho_S) = \sum\limits_k \bigg\{ \gamma_k^- \mathcal{D}[L_k] + \gamma_k^+ \mathcal{D}[L_k^\dagger]\bigg\},
\end{equation}
where $\mathcal{D}[L] = L \rho_S L^\dagger - \frac{1}{2} \{L^\dagger L, \rho\}$. 
We also define  the jump coefficients  $\gamma_k^- = |g_k|^2 \langle A_k A_k^\dagger \rangle_\text{th}$ and $\gamma_k^+ = |g_k|^2 \langle A_k^\dagger A_k \rangle_\text{th}$, with $\langle \ldots \rangle_\text{th} = \tr\big\{ (\ldots) \rho_A^\text{th}\big\}$. 
As shown e.g. in Ref.~\cite{Breuer2007}, since the $A_k$ are eigenoperators, these coefficients satisfy detailed balance $\gamma_k^+/\gamma_k^- = e^{-\beta \omega_k}$.
As for the new coherent contribution in Eq.~(\ref{G}), we now find
\begin{equation}\label{G_eigenop}
G = \sum\limits_k  \bigg\{ g_k  \langle A_k \rangle_\chi L_k^\dagger + g_k^* \langle A_k^\dagger \rangle_\chi L_k  \bigg\}, 
\end{equation}
where $\langle \ldots \rangle_\chi = \tr\big\{ (\ldots) \chi \big\}$ means an average over the coherent part $\chi$ of the ancillae. 

%
%
%
%
{\bf \emph{Qubit example - }}
As an illustrative example, suppose both system and ancillae are resonant qubits with $H_{S(A)} = \frac{\Omega}{2} \sigma_z^{S(A)}$ and $V_{SA} = g (\sigma_+^S \sigma_-^A + \sigma_-^S \sigma_+^A)$. 
Moreover, we take $\chi_A = |0\rangle\langle 1| + |1\rangle\langle 0|$, so that Eq.~(\ref{D_eigenop}) reduces  to the simple amplitude damping dissipator $D(\rho_S) = \sum_{j=\pm}\gamma^j \mathcal{D}[\sigma_j^S]$
, whereas the coherent contribution in Eq.~(\ref{G_eigenop}) goes to $G = g \sigma_x^S$.
The dynamics of the system will then mimic that of a two-level atom driven by classical light, with $D(\rho_S)$ representing the incoherent emission or absorption of radiation and $G$ a coherent driving term. 

{\bf \emph{Modified first law - }}
Collisional models enable the unambiguous distinctions between heat and work, which is in general not the case~\cite{Alicki1979}, due to the full access to the global dynamics offered by such approaches~\cite{Strasberg2016,DeChiara2018}. 
In particular, Eq.~(\ref{M}) was derived under the assumption of strong energy conservation, so that {\color{black} no work by an external agent is required to perform the unitary.}
Any energy changes in the system are thus solely due to energy leaving or entering the ancillae.

The evolution of $\langle H_S\rangle$ is easily evaluated as 
\begin{equation}\label{energy_balance_MEq}
{d \langle H_S \rangle}/{d t} = i \lambda \langle [G,H_S] \rangle + \tr\left[ H_S D(\rho_S)\right].
\end{equation}
{\color{black}The basic structure of these two terms is clearly different.}
The second term represents the typical \emph{incoherent} energy usually associated with heat, whereas the first represents a coherent contribution more akin to quantum mechanical work. 
Indeed, we  show below that, the first term in Eq.~(\ref{energy_balance_MEq})  satisfies the properties expected from quantum mechanical work. 
We shall thus refer to it as the \emph{coherent work}, $\dot{\mathcal{W}}_C =  i \lambda \langle [G, H_S]\rangle$. 
We also refer to the last term in Eq.~(\ref{energy_balance_MEq}) as the incoherent heat, $\dot{\mathcal{Q}}_\text{inc} = \tr\big\{ H_S D(\rho_S)\big\}$. 
{\color{black}As the ancillas are not thermal, they will act as both thermal and work reservoirs (in the sense specified in Ref.~\cite{Strasberg2016}). As a consequence, classifying their change of energy as  heat or work is prone to a certain level of ambiguity. 
For weakly coherent ancillas, however, this separation becomes unambiguous. 
}

Combining this with Eq.~(\ref{first_law})  gives the modified first law 
\begin{equation}\label{modified_first_law}
{d \langle H_S \rangle}/{d t} \equiv - \dot{Q}_A = \dot{\mathcal{W}}_C + \dot{\mathcal{Q}}_\text{inc},
\end{equation}
where $\dot{Q}_A = \lim_{\tau \to 0} Q_{A_n}/\tau$ is the change in energy of each ancilla. 
Such modified first law is one of our key results.
It reflects a \emph{transformation process}, where part of the heat flowing in or out of the ancillae is converted into a coherent energy change $\mathcal{W}_C$, with the remainder staying  as the incoherent heat $\dot{\mathcal{Q}}_\text{inc}$. 
Next, we show that this  transformation process is made possible by consuming  coherence in the ancillae.

%
%
%
%
{\bf \emph{Modified second law - }}
We now turn to the second law in Eq.~(\ref{second_law}).
All entropic quantities can be computed using perturbation theory in $\tau$, leading to results that become exact in the limit $\tau \to 0$. 
The details are given in Ref.~\cite{SupMat}. We find 
\begin{IEEEeqnarray}{rCl} 
\label{mutual_info_result}
\mathcal{I}(\rho_{SA_n}') &=& -\beta \Delta F - \Delta C_{A_n}	\\[0.2cm]
S(\rho^{\prime}_{A_n}||\rho_{A}) &=& \beta \mathcal{W}_C + \Delta C_{A_n},
\label{relative_entropy_result}
\end{IEEEeqnarray} 
where $\Delta F$ is the change in non-equilibrium free energy of the system, $F(\rho_S) = \langle H_S \rangle - T S(\rho_S)$ and $\mathcal{W}_C \simeq \dot{\mathcal{W}}_C \tau$. 
Moreover $\Delta C_{A_n}= C(\rho^{\prime}_{A_n}) - C(\rho_{A_n})$ is the change in the relative entropy of coherence~\cite{Baumgratz2014,Streltsov2016a} in the state of the ancillae with  
$\mathcal{C}(\rho_A) = S(\rho_A^d) - S(\rho_A)$,
with $\rho_A^d$ the diagonal part of $\rho_A$ in the eigenbasis of $H_A$.
If $\lambda = 0$ in Eq.~(\ref{ancilla_state}), we get $\mathcal{W}_C = \Delta C_{A_n} = 0$ (so that $\Sigma = - \beta \Delta F$). 

The positivity of the relative entropy in Eq.~(\ref{relative_entropy_result}) implies that in each system-ancilla interaction, the coherent work is always bounded by 
\begin{equation}\label{bound_entropy_of_coherence}
\beta \mathcal{W}_C \geq -\Delta C_{A_n}.
\end{equation}
This is the core result of our investigation: It shows that the coherent work is bounded by the loss of coherence in the state of the ancillae, which needs to be consumed in order to enable the transformation process described in Eq.~(\ref{modified_first_law}). Coherence can, in this case, therefore be interpreted as a thermodynamic resource, which must be used to convert disordered energy in the ancillae  into an ordered type of energy usable for the system. 

{\color{black}On a more general level, the resource in question here is the athermality of $\rho_A$ \cite{Brandao2013, Brandao2015,Lostaglio2015} (i.e., its non-passive character \cite{Uzdin2018}).
However, the specifics of how this resource can be extracted (which requires knowledge of the operator $G$) and, most importantly, into \emph{what} it can be converted too, will depend on the form of $\rho_A$.
This argument can be further strengthened by studying the ergotropy \cite{Allahverdyan2004} in the state~(\ref{ancilla_state}). This is defined as the maximum amount of work extractable from $\rho_A$.
As we show in~\cite{SupMat}, for  weakly coherent states it follows that $\mathcal{W} = T C(\rho_A)$. 
This provides additional physical grounds to the bound in Eq.~(\ref{bound_entropy_of_coherence}): the optimal process for extracting coherent work is when the ancillas lose all their coherence, so that $-\Delta C(\rho_A) = C(\rho_A)$. 
}

Inserting Eqs.~(\ref{mutual_info_result}) and (\ref{relative_entropy_result}) in Eq.~(\ref{second_law}) and taking the limit $\tau \to 0$, one finds that the entropy production rate $\Pi = \lim_{\tau \to 0} \Sigma_n/\tau$ can be expressed as
\begin{equation}\label{modified_second_law}
\Pi = \beta \left(\dot{\mathcal{W}}_C - \dot{F}\right) = \dot{S}(\rho_S) - \beta \dot{\mathcal{Q}}_\text{inc}. 
\end{equation}
This equation embodies a modified second law of thermodynamics in the presence of weak coherences.It is structurally identical to the classical second law \cite{Fermi1956}, but with the coherent work $\mathcal{W}_C$ instead. 
The positivity of $\Pi$ sets the bound $\dot{\mathcal{W}}_C \geq \dot{F}$ that, albeit looser than the one in Eq.~(\ref{bound_entropy_of_coherence}), has the advantage of depend solely on system-related quantities. 

%
%
%
%

\begin{figure}
\centering
\includegraphics[width=0.45\textwidth]{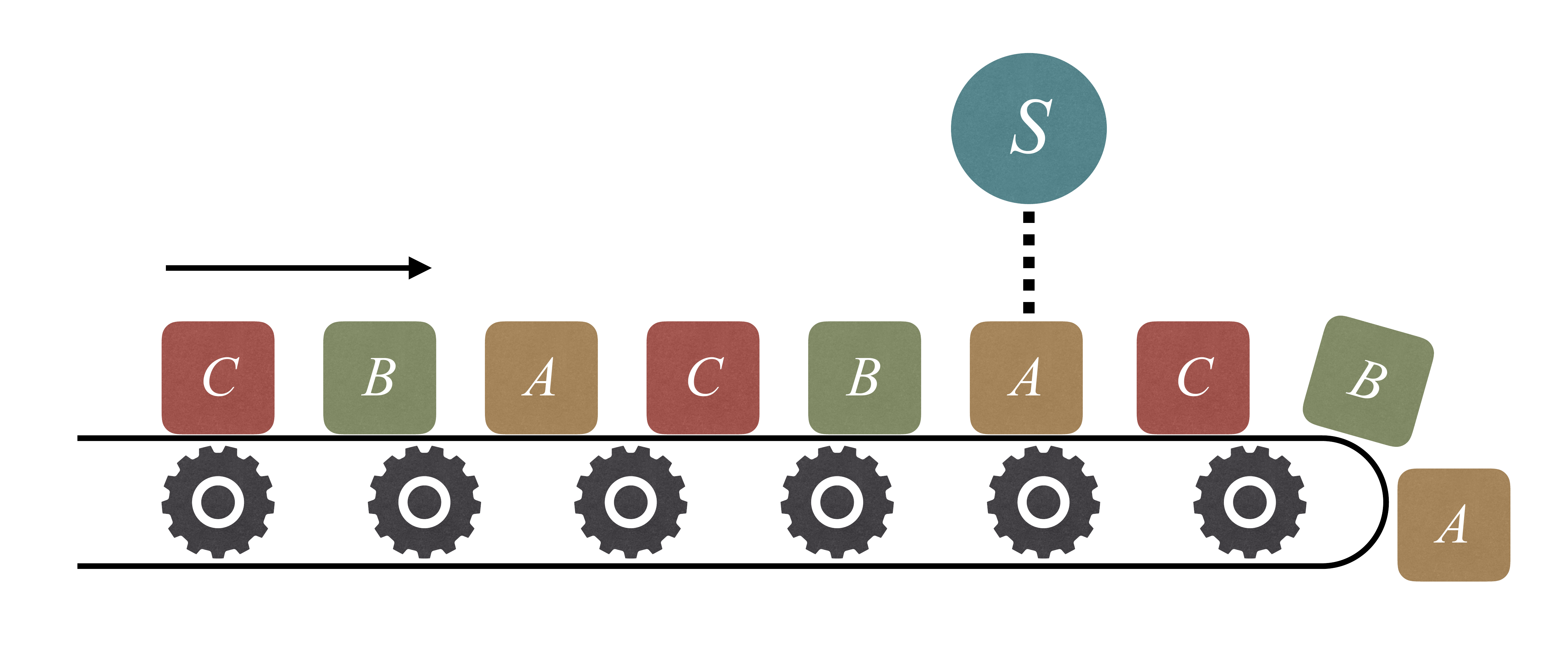}
\caption{\label{fig:drawing_multibath}Example of a collisional model where the system interacts with multiple species of ancillae.}
\end{figure}

{\bf \emph{Extension to multiple environments - }}
An extremely powerful feature of collisional models is the ability to describe systems coupled to multiple baths. 
The typical idea is represented in Fig.~\ref{fig:drawing_multibath}.
The system is placed to interact with multiple species of ancillae, with each species being independent and identically prepared in states $\rho_A$, $\rho_B$, $\rho_C$, etc. 
This can be used to model non-equilibrium steady-states, e.g. of systems coupled to multiple baths. 
In the stroboscopic scenario the state of the system will be constantly bouncing back and forth with each interaction, even in the long-time limit. 
But  the stroboscopic state after sequences of repeated interactions with the ancillae will in general converge to a steady-state. 

The remarkable feature of this construction is that  the contributions from each species become \emph{additive} in the continuous-time limit, in contrast to models where the bath is constantly coupled to the system  \cite{Mitchison2018}. 
We assume that each interaction lasts for a time  $\tau/m$, where $m$ is the number of ancilla species (e.g. $m = 3$ in Fig.~\ref{fig:drawing_multibath}).
Moreover, let  $i = A, B, C, \ldots, m$ label the different species.
To obtain a well behaved continuous-time limit, one must rescale the interaction potential $V_{Si}$ with each species [Eq.~(\ref{total_Hamiltonian})]  by $m/\sqrt{\tau}$, while keeping the coherent terms in Eq.~(\ref{ancilla_state}) proportional to $\sqrt{\tau}$. 
Using this recipe we find the master equation
\begin{equation}
\dot{\rho}= - i \big[H_S + \sum_{j} \lambda_j G_{Sj}, \rho_S\big] + \sum\limits_{j} D_j(\rho_S), 
\end{equation}
where the sums are over the various species  involved, while $G_{Sj}$ and $D_j$ are \emph{exactly} the same as those given in Eqs.~(\ref{D}) and (\ref{G}). 
This is extremely useful, as it provides a recipe to construct complex master equations, with non-trivial steady-states, from fundamental underlying building blocks.

This approach translates neatly into the first and second laws of thermodynamics, which now become
\begin{equation}\label{modified_first_law_multi_bath}
{d \langle H_S \rangle}/{d t} \equiv - \sum\limits_{j}\dot{Q}_j = \sum\limits_{j} \left(\dot{\mathcal{W}}_{C}^j + \dot{\mathcal{Q}}_\text{inc}^j \right),
\end{equation}
and 
\begin{equation}\label{modified_second_law_multi_bath}
\Pi =  \dot{S}(\rho_S) - \sum\limits_j \beta_j  \dot{\mathcal{Q}}_\text{inc}^j,
\end{equation}
where $\beta_j$ is the inverse temperature of species $j$. Both have the same structure as the usual first and second laws for systems coupled to multiple environments.

%
%
%
%
{\bf \emph{Conclusions - }} We have introduced a scenario beyond the standard system plus thermal-bath, for which operationally useful thermodynamic laws can be constructed. 
The key feature of our scenario is the use of weakly coherent states. For strong system-ancilla interactions, even weak coherences already lead to a non-trivial contribution. This leads to a modified continuous-time Lindblad master equation that encompasses a non-trivial coherent term giving rise to an effective work contribution to the energetics of the open system, although no external work is exerted at the global level. Incoherent (thermal) energy provided by the environment is catalyzed into work-like terms for the system to use by the (weak) coherence with which the former is endowed. 

We believe that this analysis thus provides a striking example of the resource-like role that coherence can play in non-equilibrium thermodynamic processes~\cite{Santos2017}.   
This could find applications, for instance, in the design of heat engines mixing classical and quantum resources. 


{\it Acknowledgements.--}
The authors acknowledge fruitful discussions with E. Lutz, A. C. Michels, J. P. Santos. FLSR acknowledges support from the Brazilian funding agency CNPq. GTL acknowledges the S\~ao Paulo Research Foundation (FAPESP) under grant numbers 2018/12813-0, 2017/50304-7. MP is supported by the EU Collaborative project TEQ (grant agreement 766900), the DfE-SFI Investigator Programme (grant 15/IA/2864), COST Action CA15220, the Royal Society Wolfson Research Fellowship ((RSWF\textbackslash R3\textbackslash183013), and the Leverhulme Trust Research Project Grant (grant nr.~RGP-2018-266). GTL and MP are grateful to the SPRINT programme supported by FAPESP and Queen's University Belfast.

\bibliography{library}

\begin{thebibliography}{52}%
\makeatletter
\providecommand \@ifxundefined [1]{%
 \@ifx{#1\undefined}
}%
\providecommand \@ifnum [1]{%
 \ifnum #1\expandafter \@firstoftwo
 \else \expandafter \@secondoftwo
 \fi
}%
\providecommand \@ifx [1]{%
 \ifx #1\expandafter \@firstoftwo
 \else \expandafter \@secondoftwo
 \fi
}%
\providecommand \natexlab [1]{#1}%
\providecommand \enquote  [1]{``#1''}%
\providecommand \bibnamefont  [1]{#1}%
\providecommand \bibfnamefont [1]{#1}%
\providecommand \citenamefont [1]{#1}%
\providecommand \href@noop [0]{\@secondoftwo}%
\providecommand \href [0]{\begingroup \@sanitize@url \@href}%
\providecommand \@href[1]{\@@startlink{#1}\@@href}%
\providecommand \@@href[1]{\endgroup#1\@@endlink}%
\providecommand \@sanitize@url [0]{\catcode `\\12\catcode `\$12\catcode
  `\&12\catcode `\#12\catcode `\^12\catcode `\_12\catcode `\%12\relax}%
\providecommand \@@startlink[1]{}%
\providecommand \@@endlink[0]{}%
\providecommand \url  [0]{\begingroup\@sanitize@url \@url }%
\providecommand \@url [1]{\endgroup\@href {#1}{\urlprefix }}%
\providecommand \urlprefix  [0]{URL }%
\providecommand \Eprint [0]{\href }%
\providecommand \doibase [0]{http://dx.doi.org/}%
\providecommand \selectlanguage [0]{\@gobble}%
\providecommand \bibinfo  [0]{\@secondoftwo}%
\providecommand \bibfield  [0]{\@secondoftwo}%
\providecommand \translation [1]{[#1]}%
\providecommand \BibitemOpen [0]{}%
\providecommand \bibitemStop [0]{}%
\providecommand \bibitemNoStop [0]{.\EOS\space}%
\providecommand \EOS [0]{\spacefactor3000\relax}%
\providecommand \BibitemShut  [1]{\csname bibitem#1\endcsname}%
\let\auto@bib@innerbib\@empty
\bibitem [{\citenamefont {Scully}\ \emph {et~al.}(2007)\citenamefont {Scully},
  \citenamefont {Zubairy},\ and\ \citenamefont {Agarwal}}]{Scully2007}%
  \BibitemOpen
  \bibfield  {author} {\bibinfo {author} {\bibfnamefont {M.~O.}\ \bibnamefont
  {Scully}}, \bibinfo {author} {\bibfnamefont {M.~S.}\ \bibnamefont {Zubairy}},
  \ and\ \bibinfo {author} {\bibfnamefont {G.~S.}\ \bibnamefont {Agarwal}},\
  }\href {\doibase 10.1126/science.1078955} {\bibfield  {journal} {\bibinfo
  {journal} {Science}\ }\textbf {\bibinfo {volume} {862}},\ \bibinfo {pages}
  {862} (\bibinfo {year} {2007})}\BibitemShut {NoStop}%
\bibitem [{\citenamefont {Dillenschneider}\ and\ \citenamefont
  {Lutz}(2009)}]{Dillenschneider2009a}%
  \BibitemOpen
  \bibfield  {author} {\bibinfo {author} {\bibfnamefont {R.}~\bibnamefont
  {Dillenschneider}}\ and\ \bibinfo {author} {\bibfnamefont {E.}~\bibnamefont
  {Lutz}},\ }\href {\doibase 10.1209/0295-5075/88/50003} {\bibfield  {journal}
  {\bibinfo  {journal} {European Physics Letters}\ }\textbf {\bibinfo {volume}
  {88}},\ \bibinfo {pages} {50003} (\bibinfo {year} {2009})}\BibitemShut
  {NoStop}%
\bibitem [{\citenamefont {Gardas}\ and\ \citenamefont
  {Deffner}(2015)}]{Gardas2015}%
  \BibitemOpen
  \bibfield  {author} {\bibinfo {author} {\bibfnamefont {B.}~\bibnamefont
  {Gardas}}\ and\ \bibinfo {author} {\bibfnamefont {S.}~\bibnamefont
  {Deffner}},\ }\href {\doibase 10.1103/PhysRevE.92.042126} {\bibfield
  {journal} {\bibinfo  {journal} {Physical Review E}\ }\textbf {\bibinfo
  {volume} {92}},\ \bibinfo {pages} {042126} (\bibinfo {year}
  {2015})}\BibitemShut {NoStop}%
\bibitem [{\citenamefont {Manzano}\ \emph {et~al.}(2016)\citenamefont
  {Manzano}, \citenamefont {Galve}, \citenamefont {Zambrini},\ and\
  \citenamefont {Parrondo}}]{Manzano2016}%
  \BibitemOpen
  \bibfield  {author} {\bibinfo {author} {\bibfnamefont {G.}~\bibnamefont
  {Manzano}}, \bibinfo {author} {\bibfnamefont {F.}~\bibnamefont {Galve}},
  \bibinfo {author} {\bibfnamefont {R.}~\bibnamefont {Zambrini}}, \ and\
  \bibinfo {author} {\bibfnamefont {J.~M.~R.}\ \bibnamefont {Parrondo}},\
  }\href {\doibase 10.1103/PhysRevE.93.052120} {\bibfield  {journal} {\bibinfo
  {journal} {Physical Review E}\ }\textbf {\bibinfo {volume} {93}},\ \bibinfo
  {pages} {052120} (\bibinfo {year} {2016})}\BibitemShut {NoStop}%
\bibitem [{\citenamefont {Manzano}(2018)}]{Manzano2018b}%
  \BibitemOpen
  \bibfield  {author} {\bibinfo {author} {\bibfnamefont {G.}~\bibnamefont
  {Manzano}},\ }\href {\doibase 10.1103/PhysRevE.98.042123} {\bibfield
  {journal} {\bibinfo  {journal} {Physical Review E}\ }\textbf {\bibinfo
  {volume} {98}},\ \bibinfo {pages} {042123} (\bibinfo {year} {2018})},\
  \Eprint {http://arxiv.org/abs/1806.07448} {arXiv:1806.07448} \BibitemShut
  {NoStop}%
\bibitem [{\citenamefont {Rossnagel}\ \emph {et~al.}(2014)\citenamefont
  {Rossnagel}, \citenamefont {Abah}, \citenamefont {Schmidt-Kaler},
  \citenamefont {Singer},\ and\ \citenamefont {Lutz}}]{Ronagel2014}%
  \BibitemOpen
  \bibfield  {author} {\bibinfo {author} {\bibfnamefont {J.}~\bibnamefont
  {Rossnagel}}, \bibinfo {author} {\bibfnamefont {O.}~\bibnamefont {Abah}},
  \bibinfo {author} {\bibfnamefont {F.}~\bibnamefont {Schmidt-Kaler}}, \bibinfo
  {author} {\bibfnamefont {K.}~\bibnamefont {Singer}}, \ and\ \bibinfo {author}
  {\bibfnamefont {E.}~\bibnamefont {Lutz}},\ }\href {\doibase
  10.1103/PhysRevLett.112.030602} {\bibfield  {journal} {\bibinfo  {journal}
  {Physical Review Letters}\ }\textbf {\bibinfo {volume} {112}},\ \bibinfo
  {pages} {030602} (\bibinfo {year} {2014})},\ \Eprint
  {http://arxiv.org/abs/1308.5935} {arXiv:1308.5935} \BibitemShut {NoStop}%
\bibitem [{\citenamefont {Manzano}\ \emph {et~al.}(2019)\citenamefont
  {Manzano}, \citenamefont {Silva},\ and\ \citenamefont
  {Parrondo}}]{Manzano2019}%
  \BibitemOpen
  \bibfield  {author} {\bibinfo {author} {\bibfnamefont {G.}~\bibnamefont
  {Manzano}}, \bibinfo {author} {\bibfnamefont {R.}~\bibnamefont {Silva}}, \
  and\ \bibinfo {author} {\bibfnamefont {J.~M.~R.}\ \bibnamefont {Parrondo}},\
  }\href {http://arxiv.org/abs/1709.00231} {\bibfield  {journal} {\bibinfo
  {journal} {Physical Review E}\ }\textbf {\bibinfo {volume} {99}},\ \bibinfo
  {pages} {042135} (\bibinfo {year} {2019})},\ \Eprint
  {http://arxiv.org/abs/1709.00231} {arXiv:1709.00231} \BibitemShut {NoStop}%
\bibitem [{\citenamefont {Ptaszynski}\ and\ \citenamefont
  {Esposito}(2019{\natexlab{a}})}]{Ptaszynski2019a}%
  \BibitemOpen
  \bibfield  {author} {\bibinfo {author} {\bibfnamefont {K.}~\bibnamefont
  {Ptaszynski}}\ and\ \bibinfo {author} {\bibfnamefont {M.}~\bibnamefont
  {Esposito}},\ }\href {\doibase 10.1103/PhysRevLett.122.150603} {\bibfield
  {journal} {\bibinfo  {journal} {Physical Review Letters}\ }\textbf {\bibinfo
  {volume} {122}},\ \bibinfo {pages} {150603} (\bibinfo {year}
  {2019}{\natexlab{a}})},\ \Eprint {http://arxiv.org/abs/1901.01093}
  {arXiv:1901.01093} \BibitemShut {NoStop}%
\bibitem [{\citenamefont {Holmes}\ \emph {et~al.}(2018)\citenamefont {Holmes},
  \citenamefont {Weidt}, \citenamefont {Jennings}, \citenamefont {Anders},\
  and\ \citenamefont {Mintert}}]{Holmes2018a}%
  \BibitemOpen
  \bibfield  {author} {\bibinfo {author} {\bibfnamefont {Z.}~\bibnamefont
  {Holmes}}, \bibinfo {author} {\bibfnamefont {S.}~\bibnamefont {Weidt}},
  \bibinfo {author} {\bibfnamefont {D.}~\bibnamefont {Jennings}}, \bibinfo
  {author} {\bibfnamefont {J.}~\bibnamefont {Anders}}, \ and\ \bibinfo {author}
  {\bibfnamefont {F.}~\bibnamefont {Mintert}},\ }\href
  {http://arxiv.org/abs/1806.11256} {\bibfield  {journal} {\bibinfo  {journal}
  {Quantum}\ }\textbf {\bibinfo {volume} {3}},\ \bibinfo {pages} {124}
  (\bibinfo {year} {2018})},\ \Eprint {http://arxiv.org/abs/1806.11256}
  {arXiv:1806.11256} \BibitemShut {NoStop}%
\bibitem [{\citenamefont {Korzekwa}\ \emph {et~al.}(2016)\citenamefont
  {Korzekwa}, \citenamefont {Lostaglio}, \citenamefont {Oppenheim},\ and\
  \citenamefont {Jennings}}]{Korzekwa2016}%
  \BibitemOpen
  \bibfield  {author} {\bibinfo {author} {\bibfnamefont {K.}~\bibnamefont
  {Korzekwa}}, \bibinfo {author} {\bibfnamefont {M.}~\bibnamefont {Lostaglio}},
  \bibinfo {author} {\bibfnamefont {J.}~\bibnamefont {Oppenheim}}, \ and\
  \bibinfo {author} {\bibfnamefont {D.}~\bibnamefont {Jennings}},\ }\href@noop
  {} {\bibfield  {journal} {\bibinfo  {journal} {New Journal of Physics}\
  }\textbf {\bibinfo {volume} {18}},\ \bibinfo {pages} {023045} (\bibinfo
  {year} {2016})}\BibitemShut {NoStop}%
\bibitem [{\citenamefont {B{\"{a}}umer}\ \emph {et~al.}(2019)\citenamefont
  {B{\"{a}}umer}, \citenamefont {Lostaglio}, \citenamefont {Perarnau-Llobet},\
  and\ \citenamefont {Sampaio}}]{Baumer2018}%
  \BibitemOpen
  \bibfield  {author} {\bibinfo {author} {\bibfnamefont {E.}~\bibnamefont
  {B{\"{a}}umer}}, \bibinfo {author} {\bibfnamefont {M.}~\bibnamefont
  {Lostaglio}}, \bibinfo {author} {\bibfnamefont {M.}~\bibnamefont
  {Perarnau-Llobet}}, \ and\ \bibinfo {author} {\bibfnamefont {R.}~\bibnamefont
  {Sampaio}},\ }in\ \href {http://arxiv.org/abs/1805.10096} {\emph {\bibinfo
  {booktitle} {Thermodynamics in the quantum regime - Fundamental Theories of
  Physics}}},\ \bibinfo {editor} {edited by\ \bibinfo {editor} {\bibfnamefont
  {F.}~\bibnamefont {Binder}}, \bibinfo {editor} {\bibfnamefont
  {L.}~\bibnamefont {Correa}}, \bibinfo {editor} {\bibfnamefont
  {C.}~\bibnamefont {Gogolin}}, \bibinfo {editor} {\bibfnamefont
  {J.}~\bibnamefont {Anders}}, \ and\ \bibinfo {editor} {\bibfnamefont
  {G.}~\bibnamefont {Adesso}}}\ (\bibinfo  {publisher} {Springer},\ \bibinfo
  {year} {2019})\ p.\ \bibinfo {pages} {195},\ \Eprint
  {http://arxiv.org/abs/1805.10096} {arXiv:1805.10096} \BibitemShut {NoStop}%
\bibitem [{\citenamefont {Micadei}\ \emph {et~al.}(2019)\citenamefont
  {Micadei}, \citenamefont {Peterson}, \citenamefont {Souza}, \citenamefont
  {Sarthour}, \citenamefont {Oliveira}, \citenamefont {Landi}, \citenamefont
  {Batalh{\~{a}}o}, \citenamefont {Serra},\ and\ \citenamefont
  {Lutz}}]{Micadei2017}%
  \BibitemOpen
  \bibfield  {author} {\bibinfo {author} {\bibfnamefont {K.}~\bibnamefont
  {Micadei}}, \bibinfo {author} {\bibfnamefont {J.~P.~S.}\ \bibnamefont
  {Peterson}}, \bibinfo {author} {\bibfnamefont {A.~M.}\ \bibnamefont {Souza}},
  \bibinfo {author} {\bibfnamefont {R.~S.}\ \bibnamefont {Sarthour}}, \bibinfo
  {author} {\bibfnamefont {I.~S.}\ \bibnamefont {Oliveira}}, \bibinfo {author}
  {\bibfnamefont {G.~T.}\ \bibnamefont {Landi}}, \bibinfo {author}
  {\bibfnamefont {T.~B.}\ \bibnamefont {Batalh{\~{a}}o}}, \bibinfo {author}
  {\bibfnamefont {R.~M.}\ \bibnamefont {Serra}}, \ and\ \bibinfo {author}
  {\bibfnamefont {E.}~\bibnamefont {Lutz}},\ }\href {\doibase
  10.1038/s41467-019-10333-7} {\bibfield  {journal} {\bibinfo  {journal}
  {Nature Communications}\ }\textbf {\bibinfo {volume} {10}},\ \bibinfo {pages}
  {2456} (\bibinfo {year} {2019})},\ \Eprint {http://arxiv.org/abs/1711.03323}
  {arXiv:1711.03323} \BibitemShut {NoStop}%
\bibitem [{\citenamefont {Klaers}\ \emph {et~al.}(2017)\citenamefont {Klaers},
  \citenamefont {Faelt}, \citenamefont {Imamoglu},\ and\ \citenamefont
  {Togan}}]{Klaers2017a}%
  \BibitemOpen
  \bibfield  {author} {\bibinfo {author} {\bibfnamefont {J.}~\bibnamefont
  {Klaers}}, \bibinfo {author} {\bibfnamefont {S.}~\bibnamefont {Faelt}},
  \bibinfo {author} {\bibfnamefont {A.}~\bibnamefont {Imamoglu}}, \ and\
  \bibinfo {author} {\bibfnamefont {E.}~\bibnamefont {Togan}},\ }\href
  {\doibase 10.1103/PhysRevX.7.031044} {\bibfield  {journal} {\bibinfo
  {journal} {Physical Review X}\ }\textbf {\bibinfo {volume} {7}},\ \bibinfo
  {pages} {031044} (\bibinfo {year} {2017})},\ \Eprint
  {http://arxiv.org/abs/1703.10024} {arXiv:1703.10024} \BibitemShut {NoStop}%
\bibitem [{\citenamefont {Scarani}\ \emph {et~al.}(2002)\citenamefont
  {Scarani}, \citenamefont {Ziman}, \citenamefont {{\v{S}}telmachovi{\v{c}}},
  \citenamefont {Gisin}, \citenamefont {Bu{\v{z}}ek},\ and\ \citenamefont
  {Bu{\v{z}}ek}}]{Scarani2002}%
  \BibitemOpen
  \bibfield  {author} {\bibinfo {author} {\bibfnamefont {V.}~\bibnamefont
  {Scarani}}, \bibinfo {author} {\bibfnamefont {M.}~\bibnamefont {Ziman}},
  \bibinfo {author} {\bibfnamefont {P.}~\bibnamefont
  {{\v{S}}telmachovi{\v{c}}}}, \bibinfo {author} {\bibfnamefont
  {N.}~\bibnamefont {Gisin}}, \bibinfo {author} {\bibfnamefont
  {V.}~\bibnamefont {Bu{\v{z}}ek}}, \ and\ \bibinfo {author} {\bibfnamefont
  {V.}~\bibnamefont {Bu{\v{z}}ek}},\ }\href {\doibase
  10.1103/PhysRevLett.88.097905} {\bibfield  {journal} {\bibinfo  {journal}
  {Physical Review Letters}\ }\textbf {\bibinfo {volume} {88}},\ \bibinfo
  {pages} {097905} (\bibinfo {year} {2002})},\ \Eprint
  {http://arxiv.org/abs/0110088} {arXiv:0110088 [quant-ph]} \BibitemShut
  {NoStop}%
\bibitem [{\citenamefont {Ziman}\ \emph {et~al.}(2002)\citenamefont {Ziman},
  \citenamefont {{\v{S}}telmachovi{\v{c}}}, \citenamefont {Buz{\v{z}}ek},
  \citenamefont {Hillery}, \citenamefont {Scarani},\ and\ \citenamefont
  {Gisin}}]{Ziman2002}%
  \BibitemOpen
  \bibfield  {author} {\bibinfo {author} {\bibfnamefont {M.}~\bibnamefont
  {Ziman}}, \bibinfo {author} {\bibfnamefont {P.}~\bibnamefont
  {{\v{S}}telmachovi{\v{c}}}}, \bibinfo {author} {\bibfnamefont
  {V.}~\bibnamefont {Buz{\v{z}}ek}}, \bibinfo {author} {\bibfnamefont
  {M.}~\bibnamefont {Hillery}}, \bibinfo {author} {\bibfnamefont
  {V.}~\bibnamefont {Scarani}}, \ and\ \bibinfo {author} {\bibfnamefont
  {N.}~\bibnamefont {Gisin}},\ }\href {\doibase 10.1103/PhysRevA.65.042105}
  {\bibfield  {journal} {\bibinfo  {journal} {Physical Review A. Atomic,
  Molecular, and Optical Physics}\ }\textbf {\bibinfo {volume} {65}},\ \bibinfo
  {pages} {042105} (\bibinfo {year} {2002})}\BibitemShut {NoStop}%
\bibitem [{\citenamefont {Karevski}\ and\ \citenamefont
  {Platini}(2009)}]{Karevski2009}%
  \BibitemOpen
  \bibfield  {author} {\bibinfo {author} {\bibfnamefont {D.}~\bibnamefont
  {Karevski}}\ and\ \bibinfo {author} {\bibfnamefont {T.}~\bibnamefont
  {Platini}},\ }\href {\doibase 10.1103/PhysRevLett.102.207207} {\bibfield
  {journal} {\bibinfo  {journal} {Physical Review Letters}\ }\textbf {\bibinfo
  {volume} {102}},\ \bibinfo {pages} {207207} (\bibinfo {year} {2009})},\
  \Eprint {http://arxiv.org/abs/0904.3527} {arXiv:0904.3527} \BibitemShut
  {NoStop}%
\bibitem [{\citenamefont {Giovannetti}\ and\ \citenamefont
  {Palma}(2012)}]{Giovannetti2012}%
  \BibitemOpen
  \bibfield  {author} {\bibinfo {author} {\bibfnamefont {V.}~\bibnamefont
  {Giovannetti}}\ and\ \bibinfo {author} {\bibfnamefont {G.~M.}\ \bibnamefont
  {Palma}},\ }\href {\doibase 10.1103/PhysRevLett.108.040401} {\bibfield
  {journal} {\bibinfo  {journal} {Physical Review Letters}\ }\textbf {\bibinfo
  {volume} {108}},\ \bibinfo {pages} {040401} (\bibinfo {year}
  {2012})}\BibitemShut {NoStop}%
\bibitem [{\citenamefont {Landi}\ \emph {et~al.}(2014)\citenamefont {Landi},
  \citenamefont {Novais}, \citenamefont {de~Oliveira},\ and\ \citenamefont
  {Karevski}}]{Landi2014b}%
  \BibitemOpen
  \bibfield  {author} {\bibinfo {author} {\bibfnamefont {G.~T.}\ \bibnamefont
  {Landi}}, \bibinfo {author} {\bibfnamefont {E.}~\bibnamefont {Novais}},
  \bibinfo {author} {\bibfnamefont {M.~J.}\ \bibnamefont {de~Oliveira}}, \ and\
  \bibinfo {author} {\bibfnamefont {D.}~\bibnamefont {Karevski}},\ }\href
  {\doibase 10.1103/PhysRevE.90.042142} {\bibfield  {journal} {\bibinfo
  {journal} {Physical Review E}\ }\textbf {\bibinfo {volume} {90}},\ \bibinfo
  {pages} {042142} (\bibinfo {year} {2014})}\BibitemShut {NoStop}%
\bibitem [{\citenamefont {Strasberg}\ \emph {et~al.}(2017)\citenamefont
  {Strasberg}, \citenamefont {Schaller}, \citenamefont {Brandes},\ and\
  \citenamefont {Esposito}}]{Strasberg2016}%
  \BibitemOpen
  \bibfield  {author} {\bibinfo {author} {\bibfnamefont {P.}~\bibnamefont
  {Strasberg}}, \bibinfo {author} {\bibfnamefont {G.}~\bibnamefont {Schaller}},
  \bibinfo {author} {\bibfnamefont {T.}~\bibnamefont {Brandes}}, \ and\
  \bibinfo {author} {\bibfnamefont {M.}~\bibnamefont {Esposito}},\ }\href
  {\doibase 10.1103/PhysRevX.7.021003} {\bibfield  {journal} {\bibinfo
  {journal} {Physical Review X}\ }\textbf {\bibinfo {volume} {7}},\ \bibinfo
  {pages} {021003} (\bibinfo {year} {2017})},\ \Eprint
  {http://arxiv.org/abs/1610.01829} {arXiv:1610.01829} \BibitemShut {NoStop}%
\bibitem [{\citenamefont {Barra}(2015)}]{Barra2015}%
  \BibitemOpen
  \bibfield  {author} {\bibinfo {author} {\bibfnamefont {F.}~\bibnamefont
  {Barra}},\ }\href {\doibase 10.1038/srep14873} {\bibfield  {journal}
  {\bibinfo  {journal} {Scientific Reports}\ }\textbf {\bibinfo {volume} {5}},\
  \bibinfo {pages} {14873} (\bibinfo {year} {2015})},\ \Eprint
  {http://arxiv.org/abs/1509.04223} {arXiv:1509.04223} \BibitemShut {NoStop}%
\bibitem [{\citenamefont {Pereira}(2018)}]{Pereira2018}%
  \BibitemOpen
  \bibfield  {author} {\bibinfo {author} {\bibfnamefont {E.}~\bibnamefont
  {Pereira}},\ }\href {\doibase 10.1103/PhysRevE.97.022115} {\bibfield
  {journal} {\bibinfo  {journal} {Physical Review E}\ }\textbf {\bibinfo
  {volume} {97}},\ \bibinfo {pages} {022115} (\bibinfo {year}
  {2018})}\BibitemShut {NoStop}%
\bibitem [{\citenamefont {Lorenzo}\ \emph
  {et~al.}(2015{\natexlab{a}})\citenamefont {Lorenzo}, \citenamefont
  {McCloskey}, \citenamefont {Ciccarello}, \citenamefont {Paternostro},\ and\
  \citenamefont {Palma}}]{Lorenzo2015}%
  \BibitemOpen
  \bibfield  {author} {\bibinfo {author} {\bibfnamefont {S.}~\bibnamefont
  {Lorenzo}}, \bibinfo {author} {\bibfnamefont {R.}~\bibnamefont {McCloskey}},
  \bibinfo {author} {\bibfnamefont {F.}~\bibnamefont {Ciccarello}}, \bibinfo
  {author} {\bibfnamefont {M.}~\bibnamefont {Paternostro}}, \ and\ \bibinfo
  {author} {\bibfnamefont {G.}~\bibnamefont {Palma}},\ }\href {\doibase
  10.1103/PhysRevLett.115.120403} {\bibfield  {journal} {\bibinfo  {journal}
  {Physical Review Letters}\ }\textbf {\bibinfo {volume} {115}},\ \bibinfo
  {pages} {120403} (\bibinfo {year} {2015}{\natexlab{a}})},\ \Eprint
  {http://arxiv.org/abs/arXiv:1503.07837v2} {arXiv:arXiv:1503.07837v2}
  \BibitemShut {NoStop}%
\bibitem [{\citenamefont {Lorenzo}\ \emph
  {et~al.}(2015{\natexlab{b}})\citenamefont {Lorenzo}, \citenamefont {Farace},
  \citenamefont {Ciccarello}, \citenamefont {Palma},\ and\ \citenamefont
  {Giovannetti}}]{Lorenzo2015b}%
  \BibitemOpen
  \bibfield  {author} {\bibinfo {author} {\bibfnamefont {S.}~\bibnamefont
  {Lorenzo}}, \bibinfo {author} {\bibfnamefont {A.}~\bibnamefont {Farace}},
  \bibinfo {author} {\bibfnamefont {F.}~\bibnamefont {Ciccarello}}, \bibinfo
  {author} {\bibfnamefont {G.~M.}\ \bibnamefont {Palma}}, \ and\ \bibinfo
  {author} {\bibfnamefont {V.}~\bibnamefont {Giovannetti}},\ }\href {\doibase
  10.1103/PhysRevA.91.022121} {\bibfield  {journal} {\bibinfo  {journal}
  {Physical Review A}\ }\textbf {\bibinfo {volume} {91}},\ \bibinfo {pages}
  {022121} (\bibinfo {year} {2015}{\natexlab{b}})}\BibitemShut {NoStop}%
\bibitem [{\citenamefont {Pezzutto}\ \emph {et~al.}(2016)\citenamefont
  {Pezzutto}, \citenamefont {Paternostro},\ and\ \citenamefont
  {Omar}}]{Pezzutto2016}%
  \BibitemOpen
  \bibfield  {author} {\bibinfo {author} {\bibfnamefont {M.}~\bibnamefont
  {Pezzutto}}, \bibinfo {author} {\bibfnamefont {M.}~\bibnamefont
  {Paternostro}}, \ and\ \bibinfo {author} {\bibfnamefont {Y.}~\bibnamefont
  {Omar}},\ }\href {\doibase 10.1088/1367-2630/18/12/123018} {\bibfield
  {journal} {\bibinfo  {journal} {New Journal of Physics}\ }\textbf {\bibinfo
  {volume} {18}},\ \bibinfo {pages} {123018} (\bibinfo {year}
  {2016})}\BibitemShut {NoStop}%
\bibitem [{\citenamefont {Pezzutto}\ \emph {et~al.}(2019)\citenamefont
  {Pezzutto}, \citenamefont {Paternostro},\ and\ \citenamefont
  {Omar}}]{Pezzutto2019}%
  \BibitemOpen
  \bibfield  {author} {\bibinfo {author} {\bibfnamefont {M.}~\bibnamefont
  {Pezzutto}}, \bibinfo {author} {\bibfnamefont {M.}~\bibnamefont
  {Paternostro}}, \ and\ \bibinfo {author} {\bibfnamefont {Y.}~\bibnamefont
  {Omar}},\ }\href {\doibase 10.1088/2058-9565/aaf5b4} {\bibfield  {journal}
  {\bibinfo  {journal} {Quantum Science and Technology}\ }\textbf {\bibinfo
  {volume} {4}},\ \bibinfo {pages} {025002} (\bibinfo {year} {2019})},\ \Eprint
  {http://arxiv.org/abs/1806.10075} {arXiv:1806.10075} \BibitemShut {NoStop}%
\bibitem [{\citenamefont {Cusumano}\ \emph {et~al.}(2018)\citenamefont
  {Cusumano}, \citenamefont {Cavina}, \citenamefont {Keck}, \citenamefont {{De
  Pasquale}},\ and\ \citenamefont {Giovannetti}}]{Cusumano2018}%
  \BibitemOpen
  \bibfield  {author} {\bibinfo {author} {\bibfnamefont {S.}~\bibnamefont
  {Cusumano}}, \bibinfo {author} {\bibfnamefont {V.}~\bibnamefont {Cavina}},
  \bibinfo {author} {\bibfnamefont {M.}~\bibnamefont {Keck}}, \bibinfo {author}
  {\bibfnamefont {A.}~\bibnamefont {{De Pasquale}}}, \ and\ \bibinfo {author}
  {\bibfnamefont {V.}~\bibnamefont {Giovannetti}},\ }\href {\doibase
  10.1103/PhysRevA.98.032119} {\bibfield  {journal} {\bibinfo  {journal}
  {Physical Review A}\ }\textbf {\bibinfo {volume} {98}},\ \bibinfo {pages}
  {032119} (\bibinfo {year} {2018})},\ \Eprint
  {http://arxiv.org/abs/arXiv:1807.04500v2} {arXiv:arXiv:1807.04500v2}
  \BibitemShut {NoStop}%
\bibitem [{\citenamefont {Englert}\ and\ \citenamefont
  {Morigi}(2002)}]{Englert2002}%
  \BibitemOpen
  \bibfield  {author} {\bibinfo {author} {\bibfnamefont {B.-G.}\ \bibnamefont
  {Englert}}\ and\ \bibinfo {author} {\bibfnamefont {G.}~\bibnamefont
  {Morigi}},\ }in\ \href@noop {} {\emph {\bibinfo {booktitle} {Coherent
  Evolution in Noisy Environments - Lecture Notes in Physics}}},\ \bibinfo
  {editor} {edited by\ \bibinfo {editor} {\bibfnamefont {A.}~\bibnamefont
  {Buchleitner}}\ and\ \bibinfo {editor} {\bibfnamefont {K.}~\bibnamefont
  {Hornberger}}}\ (\bibinfo  {publisher} {Springer},\ \bibinfo {address}
  {Berlin, Heidelberg},\ \bibinfo {year} {2002})\ p.\ \bibinfo {pages} {611},\
  \Eprint {http://arxiv.org/abs/0206116} {arXiv:0206116 [quant-ph]}
  \BibitemShut {NoStop}%
\bibitem [{\citenamefont {Doob}(1954)}]{Doob1954}%
  \BibitemOpen
  \bibfield  {author} {\bibinfo {author} {\bibfnamefont {J.~L.}\ \bibnamefont
  {Doob}},\ }in\ \href@noop {} {\emph {\bibinfo {booktitle} {Selected Papers on
  Noise and Stochastic Processes}}},\ \bibinfo {editor} {edited by\ \bibinfo
  {editor} {\bibfnamefont {N.}~\bibnamefont {Wax}}}\ (\bibinfo  {publisher}
  {Dover},\ \bibinfo {address} {New York},\ \bibinfo {year} {1954})\BibitemShut
  {NoStop}%
\bibitem [{\citenamefont {Brand{\~{a}}o}\ \emph {et~al.}(2013)\citenamefont
  {Brand{\~{a}}o}, \citenamefont {Horodecki}, \citenamefont {Oppenheim},
  \citenamefont {Renes},\ and\ \citenamefont {Spekkens}}]{Brandao2013}%
  \BibitemOpen
  \bibfield  {author} {\bibinfo {author} {\bibfnamefont {F.~G. S.~L.}\
  \bibnamefont {Brand{\~{a}}o}}, \bibinfo {author} {\bibfnamefont
  {M.}~\bibnamefont {Horodecki}}, \bibinfo {author} {\bibfnamefont
  {J.}~\bibnamefont {Oppenheim}}, \bibinfo {author} {\bibfnamefont {J.~M.}\
  \bibnamefont {Renes}}, \ and\ \bibinfo {author} {\bibfnamefont {R.~W.}\
  \bibnamefont {Spekkens}},\ }\href {\doibase 10.1103/PhysRevLett.111.250404}
  {\bibfield  {journal} {\bibinfo  {journal} {Physical Review Letters}\
  }\textbf {\bibinfo {volume} {111}},\ \bibinfo {pages} {250404} (\bibinfo
  {year} {2013})},\ \Eprint {http://arxiv.org/abs/1111.3882} {arXiv:1111.3882}
  \BibitemShut {NoStop}%
\bibitem [{\citenamefont {Brand{\~{a}}o}\ \emph {et~al.}(2015)\citenamefont
  {Brand{\~{a}}o}, \citenamefont {Horodecki}, \citenamefont {Ng}, \citenamefont
  {Oppenheim},\ and\ \citenamefont {Wehner}}]{Brandao2015}%
  \BibitemOpen
  \bibfield  {author} {\bibinfo {author} {\bibfnamefont {F.~G. S.~L.}\
  \bibnamefont {Brand{\~{a}}o}}, \bibinfo {author} {\bibfnamefont
  {M.}~\bibnamefont {Horodecki}}, \bibinfo {author} {\bibfnamefont {N.~H.~Y.}\
  \bibnamefont {Ng}}, \bibinfo {author} {\bibfnamefont {J.}~\bibnamefont
  {Oppenheim}}, \ and\ \bibinfo {author} {\bibfnamefont {S.}~\bibnamefont
  {Wehner}},\ }\href {\doibase 10.1073/pnas.1411728112} {\bibfield  {journal}
  {\bibinfo  {journal} {Proceedings of the National Academy of Sciences}\
  }\textbf {\bibinfo {volume} {112}},\ \bibinfo {pages} {3275} (\bibinfo {year}
  {2015})},\ \Eprint {http://arxiv.org/abs/1305.5278} {arXiv:1305.5278}
  \BibitemShut {NoStop}%
\bibitem [{\citenamefont {Santos}\ \emph {et~al.}(2019)\citenamefont {Santos},
  \citenamefont {C{\'{e}}leri}, \citenamefont {Landi},\ and\ \citenamefont
  {Paternostro}}]{Santos2017}%
  \BibitemOpen
  \bibfield  {author} {\bibinfo {author} {\bibfnamefont {J.~P.}\ \bibnamefont
  {Santos}}, \bibinfo {author} {\bibfnamefont {L.~C.}\ \bibnamefont
  {C{\'{e}}leri}}, \bibinfo {author} {\bibfnamefont {G.~T.}\ \bibnamefont
  {Landi}}, \ and\ \bibinfo {author} {\bibfnamefont {M.}~\bibnamefont
  {Paternostro}},\ }\href {\doibase https://doi.org/10.1038/s41534-019-0138-y}
  {\bibfield  {journal} {\bibinfo  {journal} {Nature Quantum Information
  (accepted)}\ }\textbf {\bibinfo {volume} {5}},\ \bibinfo {pages} {23}
  (\bibinfo {year} {2019})},\ \Eprint {http://arxiv.org/abs/1707.08946}
  {arXiv:1707.08946} \BibitemShut {NoStop}%
\bibitem [{\citenamefont {{De Chiara}}\ \emph {et~al.}(2018)\citenamefont {{De
  Chiara}}, \citenamefont {Landi}, \citenamefont {Hewgill}, \citenamefont
  {Reid}, \citenamefont {Ferraro}, \citenamefont {Roncaglia},\ and\
  \citenamefont {Antezza}}]{DeChiara2018}%
  \BibitemOpen
  \bibfield  {author} {\bibinfo {author} {\bibfnamefont {G.}~\bibnamefont {{De
  Chiara}}}, \bibinfo {author} {\bibfnamefont {G.}~\bibnamefont {Landi}},
  \bibinfo {author} {\bibfnamefont {A.}~\bibnamefont {Hewgill}}, \bibinfo
  {author} {\bibfnamefont {B.}~\bibnamefont {Reid}}, \bibinfo {author}
  {\bibfnamefont {A.}~\bibnamefont {Ferraro}}, \bibinfo {author} {\bibfnamefont
  {A.~J.}\ \bibnamefont {Roncaglia}}, \ and\ \bibinfo {author} {\bibfnamefont
  {M.}~\bibnamefont {Antezza}},\ }\href {\doibase
  https://doi.org/10.1088/1367-2630/aaecee} {\bibfield  {journal} {\bibinfo
  {journal} {New Journal of Physics}\ }\textbf {\bibinfo {volume} {20}},\
  \bibinfo {pages} {113024} (\bibinfo {year} {2018})},\ \Eprint
  {http://arxiv.org/abs/1808.10450} {arXiv:1808.10450} \BibitemShut {NoStop}%
\bibitem [{\citenamefont {Manzano}\ \emph {et~al.}(2018)\citenamefont
  {Manzano}, \citenamefont {Horowitz},\ and\ \citenamefont
  {Parrondo}}]{Manzano2017a}%
  \BibitemOpen
  \bibfield  {author} {\bibinfo {author} {\bibfnamefont {G.}~\bibnamefont
  {Manzano}}, \bibinfo {author} {\bibfnamefont {J.~M.}\ \bibnamefont
  {Horowitz}}, \ and\ \bibinfo {author} {\bibfnamefont {J.~M.~R.}\ \bibnamefont
  {Parrondo}},\ }\href {http://arxiv.org/abs/1710.00054} {\bibfield  {journal}
  {\bibinfo  {journal} {Physical Review X}\ }\textbf {\bibinfo {volume} {8}},\
  \bibinfo {pages} {031037} (\bibinfo {year} {2018})},\ \Eprint
  {http://arxiv.org/abs/1710.00054} {arXiv:1710.00054} \BibitemShut {NoStop}%
\bibitem [{\citenamefont {Reeb}\ and\ \citenamefont {Wolf}(2014)}]{Reeb2014}%
  \BibitemOpen
  \bibfield  {author} {\bibinfo {author} {\bibfnamefont {D.}~\bibnamefont
  {Reeb}}\ and\ \bibinfo {author} {\bibfnamefont {M.~M.}\ \bibnamefont
  {Wolf}},\ }\href {\doibase 10.1088/1367-2630/16/10/103011} {\bibfield
  {journal} {\bibinfo  {journal} {New Journal of Physics}\ }\textbf {\bibinfo
  {volume} {16}},\ \bibinfo {pages} {103011} (\bibinfo {year} {2014})},\
  \Eprint {http://arxiv.org/abs/1306.4352} {arXiv:1306.4352} \BibitemShut
  {NoStop}%
\bibitem [{\citenamefont {Talkner}\ \emph {et~al.}(2009)\citenamefont
  {Talkner}, \citenamefont {Campisi},\ and\ \citenamefont
  {H{\"{a}}nggi}}]{Talkner2009}%
  \BibitemOpen
  \bibfield  {author} {\bibinfo {author} {\bibfnamefont {P.}~\bibnamefont
  {Talkner}}, \bibinfo {author} {\bibfnamefont {M.}~\bibnamefont {Campisi}}, \
  and\ \bibinfo {author} {\bibfnamefont {P.}~\bibnamefont {H{\"{a}}nggi}},\
  }\href {\doibase 10.1088/1742-5468/2009/02/P02025} {\bibfield  {journal}
  {\bibinfo  {journal} {Journal of Statistical Mechanics: Theory and
  Experiment}\ }\textbf {\bibinfo {volume} {2009}},\ \bibinfo {pages} {P02025}
  (\bibinfo {year} {2009})}\BibitemShut {NoStop}%
\bibitem [{\citenamefont {Goold}\ \emph {et~al.}(2015)\citenamefont {Goold},
  \citenamefont {Paternostro},\ and\ \citenamefont {Modi}}]{Goold2014b}%
  \BibitemOpen
  \bibfield  {author} {\bibinfo {author} {\bibfnamefont {J.}~\bibnamefont
  {Goold}}, \bibinfo {author} {\bibfnamefont {M.}~\bibnamefont {Paternostro}},
  \ and\ \bibinfo {author} {\bibfnamefont {K.}~\bibnamefont {Modi}},\ }\href
  {\doibase 10.1103/PhysRevLett.114.060602} {\bibfield  {journal} {\bibinfo
  {journal} {Physical Review Letters}\ }\textbf {\bibinfo {volume} {114}},\
  \bibinfo {pages} {060602} (\bibinfo {year} {2015})}\BibitemShut {NoStop}%
\bibitem [{\citenamefont {Ptaszynski}\ and\ \citenamefont
  {Esposito}(2019{\natexlab{b}})}]{Ptaszynski2019b}%
  \BibitemOpen
  \bibfield  {author} {\bibinfo {author} {\bibfnamefont {K.}~\bibnamefont
  {Ptaszynski}}\ and\ \bibinfo {author} {\bibfnamefont {M.}~\bibnamefont
  {Esposito}},\ }\href {http://arxiv.org/abs/1905.03804} {\  (\bibinfo {year}
  {2019}{\natexlab{b}})},\ \Eprint {http://arxiv.org/abs/1905.03804}
  {arXiv:1905.03804} \BibitemShut {NoStop}%
\bibitem [{Note1()}]{Note1}%
  \BibitemOpen
  \bibinfo {note} {For instance, the white noise appearing in the Langevin
  equation of Brownian motion acts for an infinitesimal time so, in order to be
  non-trivial, it has to also be infinitely strong \cite
  {Coffey2004}.}\BibitemShut {Stop}%
\bibitem [{\citenamefont {Ciccarello}(2017)}]{Ciccarello2017}%
  \BibitemOpen
  \bibfield  {author} {\bibinfo {author} {\bibfnamefont {F.}~\bibnamefont
  {Ciccarello}},\ }\href {\doibase 10.1515/qmetro-2017-0007} {\bibfield
  {journal} {\bibinfo  {journal} {Quantum Measurements and Quantum Metrology}\
  }\textbf {\bibinfo {volume} {4}},\ \bibinfo {pages} {53} (\bibinfo {year}
  {2017})},\ \Eprint {http://arxiv.org/abs/1712.04994} {arXiv:1712.04994}
  \BibitemShut {NoStop}%
\bibitem [{Sup()}]{SupMat}%
  \BibitemOpen
  \href@noop {} {\emph {\bibinfo {title} {{See supplemental
  material}}}}\BibitemShut {NoStop}%
\bibitem [{\citenamefont {Rivas}\ and\ \citenamefont
  {Huelga}(2012)}]{Rivas2012}%
  \BibitemOpen
  \bibfield  {author} {\bibinfo {author} {\bibfnamefont {{\'{A}}.}~\bibnamefont
  {Rivas}}\ and\ \bibinfo {author} {\bibfnamefont {F.~S.}\ \bibnamefont
  {Huelga}},\ }\href@noop {} {\emph {\bibinfo {title} {{Open quantum systems:
  an introduction}}}}\ (\bibinfo  {publisher} {Springer},\ \bibinfo {address}
  {Heidelberg},\ \bibinfo {year} {2012})\BibitemShut {NoStop}%
\bibitem [{\citenamefont {Breuer}\ and\ \citenamefont
  {Petruccione}(2007)}]{Breuer2007}%
  \BibitemOpen
  \bibfield  {author} {\bibinfo {author} {\bibfnamefont {H.~P.}\ \bibnamefont
  {Breuer}}\ and\ \bibinfo {author} {\bibfnamefont {F.}~\bibnamefont
  {Petruccione}},\ }\href@noop {} {\emph {\bibinfo {title} {{The Theory of Open
  Quantum Systems}}}}\ (\bibinfo  {publisher} {Oxford University Press, USA},\
  \bibinfo {year} {2007})\ p.\ \bibinfo {pages} {636}\BibitemShut {NoStop}%
\bibitem [{\citenamefont {Alicki}(1979)}]{Alicki1979}%
  \BibitemOpen
  \bibfield  {author} {\bibinfo {author} {\bibfnamefont {R.}~\bibnamefont
  {Alicki}},\ }\href {\doibase 10.1063/1.446862} {\bibfield  {journal}
  {\bibinfo  {journal} {Journal of Physics A: Mathematical and General}\
  }\textbf {\bibinfo {volume} {12}},\ \bibinfo {pages} {L103} (\bibinfo {year}
  {1979})}\BibitemShut {NoStop}%
\bibitem [{\citenamefont {Baumgratz}\ \emph {et~al.}(2014)\citenamefont
  {Baumgratz}, \citenamefont {Cramer},\ and\ \citenamefont
  {Plenio}}]{Baumgratz2014}%
  \BibitemOpen
  \bibfield  {author} {\bibinfo {author} {\bibfnamefont {T.}~\bibnamefont
  {Baumgratz}}, \bibinfo {author} {\bibfnamefont {M.}~\bibnamefont {Cramer}}, \
  and\ \bibinfo {author} {\bibfnamefont {M.~B.}\ \bibnamefont {Plenio}},\
  }\href {\doibase 10.1103/PhysRevLett.113.140401} {\bibfield  {journal}
  {\bibinfo  {journal} {Physical Review Letters}\ }\textbf {\bibinfo {volume}
  {113}},\ \bibinfo {pages} {140401} (\bibinfo {year} {2014})},\ \Eprint
  {http://arxiv.org/abs/1311.0275} {arXiv:1311.0275} \BibitemShut {NoStop}%
\bibitem [{\citenamefont {Streltsov}\ \emph {et~al.}(2017)\citenamefont
  {Streltsov}, \citenamefont {Adesso},\ and\ \citenamefont
  {Plenio}}]{Streltsov2016a}%
  \BibitemOpen
  \bibfield  {author} {\bibinfo {author} {\bibfnamefont {A.}~\bibnamefont
  {Streltsov}}, \bibinfo {author} {\bibfnamefont {G.}~\bibnamefont {Adesso}}, \
  and\ \bibinfo {author} {\bibfnamefont {M.~B.}\ \bibnamefont {Plenio}},\
  }\href {\doibase 10.1103/RevModPhys.89.041003} {\bibfield  {journal}
  {\bibinfo  {journal} {Reviews of Modern Physics}\ }\textbf {\bibinfo {volume}
  {89}},\ \bibinfo {pages} {041003} (\bibinfo {year} {2017})},\ \Eprint
  {http://arxiv.org/abs/1609.02439} {arXiv:1609.02439} \BibitemShut {NoStop}%
\bibitem [{\citenamefont {Lostaglio}\ \emph {et~al.}(2015)\citenamefont
  {Lostaglio}, \citenamefont {Jennings},\ and\ \citenamefont
  {Rudolph}}]{Lostaglio2015}%
  \BibitemOpen
  \bibfield  {author} {\bibinfo {author} {\bibfnamefont {M.}~\bibnamefont
  {Lostaglio}}, \bibinfo {author} {\bibfnamefont {D.}~\bibnamefont {Jennings}},
  \ and\ \bibinfo {author} {\bibfnamefont {T.}~\bibnamefont {Rudolph}},\ }\href
  {\doibase 10.1038/ncomms7383} {\bibfield  {journal} {\bibinfo  {journal}
  {Nature communications}\ }\textbf {\bibinfo {volume} {6}},\ \bibinfo {pages}
  {6383} (\bibinfo {year} {2015})},\ \Eprint {http://arxiv.org/abs/1405.2188}
  {arXiv:1405.2188} \BibitemShut {NoStop}%
\bibitem [{\citenamefont {Uzdin}\ and\ \citenamefont
  {Rahav}(2018)}]{Uzdin2018}%
  \BibitemOpen
  \bibfield  {author} {\bibinfo {author} {\bibfnamefont {R.}~\bibnamefont
  {Uzdin}}\ and\ \bibinfo {author} {\bibfnamefont {S.}~\bibnamefont {Rahav}},\
  }\href {http://arxiv.org/abs/1805.00220} {\bibfield  {journal} {\bibinfo
  {journal} {Physical Review X}\ }\textbf {\bibinfo {volume} {8}},\ \bibinfo
  {pages} {021064} (\bibinfo {year} {2018})},\ \Eprint
  {http://arxiv.org/abs/1805.00220} {arXiv:1805.00220} \BibitemShut {NoStop}%
\bibitem [{\citenamefont {Allahverdyan}\ \emph {et~al.}(2004)\citenamefont
  {Allahverdyan}, \citenamefont {Balian},\ and\ \citenamefont
  {Nieuwenhuizen}}]{Allahverdyan2004}%
  \BibitemOpen
  \bibfield  {author} {\bibinfo {author} {\bibfnamefont {A.~E.}\ \bibnamefont
  {Allahverdyan}}, \bibinfo {author} {\bibfnamefont {R.}~\bibnamefont
  {Balian}}, \ and\ \bibinfo {author} {\bibfnamefont {T.~M.}\ \bibnamefont
  {Nieuwenhuizen}},\ }\href {\doibase 10.1209/epl/i2004-10101-2} {\bibfield
  {journal} {\bibinfo  {journal} {Europhysics Letters}\ }\textbf {\bibinfo
  {volume} {67}},\ \bibinfo {pages} {565} (\bibinfo {year} {2004})},\ \Eprint
  {http://arxiv.org/abs/0401574} {arXiv:0401574 [cond-mat]} \BibitemShut
  {NoStop}%
\bibitem [{\citenamefont {Fermi}(1956)}]{Fermi1956}%
  \BibitemOpen
  \bibfield  {author} {\bibinfo {author} {\bibfnamefont {E.}~\bibnamefont
  {Fermi}},\ }\href@noop {} {\emph {\bibinfo {title} {{Thermodynamics}}}}\
  (\bibinfo  {publisher} {Dover Publications Inc.},\ \bibinfo {year} {1956})\
  p.\ \bibinfo {pages} {160}\BibitemShut {NoStop}%
\bibitem [{\citenamefont {Mitchison}\ and\ \citenamefont
  {Plenio}(2018)}]{Mitchison2018}%
  \BibitemOpen
  \bibfield  {author} {\bibinfo {author} {\bibfnamefont {M.~T.}\ \bibnamefont
  {Mitchison}}\ and\ \bibinfo {author} {\bibfnamefont {M.~B.}\ \bibnamefont
  {Plenio}},\ }\href {\doibase 10.1088/1367-2630/aa9f70} {\bibfield  {journal}
  {\bibinfo  {journal} {New Journal of Physics}\ }\textbf {\bibinfo {volume}
  {20}},\ \bibinfo {pages} {033005} (\bibinfo {year} {2018})},\ \Eprint
  {http://arxiv.org/abs/1708.05574} {arXiv:1708.05574} \BibitemShut {NoStop}%
\bibitem [{\citenamefont {Coffey}\ \emph {et~al.}(2004)\citenamefont {Coffey},
  \citenamefont {Kalmykov},\ and\ \citenamefont {Waldron}}]{Coffey2004}%
  \BibitemOpen
  \bibfield  {author} {\bibinfo {author} {\bibfnamefont {W.~T.}\ \bibnamefont
  {Coffey}}, \bibinfo {author} {\bibfnamefont {Y.~P.}\ \bibnamefont
  {Kalmykov}}, \ and\ \bibinfo {author} {\bibfnamefont {J.~T.}\ \bibnamefont
  {Waldron}},\ }\href@noop {} {\emph {\bibinfo {title} {{The Langevin Equation.
  With Applications to Stochastic Problems in Physics, Chemistry and Electrical
  Engineering}}}},\ \bibinfo {edition} {2nd}\ ed.\ (\bibinfo  {publisher}
  {World Scientific Publishing Co, Pte. Ltd.},\ \bibinfo {address}
  {Singapore},\ \bibinfo {year} {2004})\ p.\ \bibinfo {pages} {678}\BibitemShut
  {NoStop}%
\bibitem [{\citenamefont {Chen}(2010)}]{Chen2010a}%
  \BibitemOpen
  \bibfield  {author} {\bibinfo {author} {\bibfnamefont {X.~Y.}\ \bibnamefont
  {Chen}},\ }\href {\doibase 10.1088/1674-1056/19/4/040308} {\bibfield
  {journal} {\bibinfo  {journal} {Chinese Physics B}\ }\textbf {\bibinfo
  {volume} {19}},\ \bibinfo {pages} {1} (\bibinfo {year} {2010})},\ \Eprint
  {http://arxiv.org/abs/0902.4733} {arXiv:0902.4733} \BibitemShut {NoStop}%
\end{thebibliography}%
\pagebreak
\widetext

\newpage 
\begin{center}
\vskip0.5cm
{\Large {\bf Supplemental Material}}
\end{center}

\bigskip
\setcounter{equation}{0}
\setcounter{figure}{0}
\setcounter{table}{0}
\setcounter{page}{1}
\renewcommand{\theequation}{S\arabic{equation}}
\renewcommand{\thefigure}{S\arabic{figure}}

In this supplemental material we provide additional details on the mathematical derivations of our most relevant results. 
In Sec.~S1 we discuss the derivation of Eq.~(\ref{M}) of the main text. 
In Sec.~S2 we discuss how to use perturbation theory to compute the von Neumann entropy and related quantities. 
These are then used in Sec.~S3 to derive Eqs. (\ref{mutual_info_result}) and (\ref{relative_entropy_result}) of the main text.

%
%
\section{S1. Derivation of Eq.~(\ref{M}) of the main text}
%
%

The derivation of Eq.~(\ref{M}) is straightforward once the basic ingredients are properly defined. 
We consider for simplicity a single system-ancilla interaction event. 
The system is prepared in an arbitrary state $\rho_S$, whereas the ancilla is prepared in the weakly coherent state $\rho_A$ in Eq.~(\ref{ancilla_state}) of the main text. 
We then apply the unitary $U_{SA}$ generated by the Hamiltonian~(\ref{total_Hamiltonian}). 
A Baker-Campbell-Haussdorf series expansion in $\tau$ leads to 
\begin{equation}\label{SM_rhoSA_prime}
\rho_{SA}' = \rho_S \rho_A - i \tau [H_{SA}, \rho_S \rho_A] - \frac{\tau^2}{2} [H_{SA}, [H_{SA},\rho_S \rho_A]].
\end{equation}
Using the specific scalings of $V_{SA}$ and $\rho_A$, and keeping only terms at most linear in $\tau$ we  get 
\begin{equation}\label{SM_unitary_expansion}
\rho_{SA}'  = \rho_S \rho_A - i \tau [H_S + H_A, \rho_S \rho_A] - i \sqrt{\tau} \; [V_{SA}, \rho_S \rho_A] - \frac{\tau}{2} [V_{SA},[V_{SA},\rho_S \rho_A^\text{th}]] + \mathcal{O}(\tau)^{3/2}.
\end{equation}
In the last term we  neglected a contribution from the coherent part $\chi_A$, as this would lead to a term at least of order $\tau^{3/2}$. 
However, in the firs two terms we kept the full $\rho_A = \rho_A^\text{th} + \sqrt{\lambda^2\tau} \chi_A$.

Eq.~(\ref{M}) of the main text now follows directly by taking the trace of Eq.~(\ref{SM_unitary_expansion}) over $A$ and assuming that $\tr_A (V_{SA} \rho_A^\text{th}) = 0$, which leads to
\begin{equation}\label{SM_rhoS_prime}{
\rho_S' = \rho_S - i \tau [H_S, \rho_S] - i \lambda \tau [G, \rho_S] + \tau D(\rho_S), 
}\end{equation}
with $D$ and $G$ given in Eqs.~(\ref{D}) and (\ref{G}) of the main text. 
Dividing both sides by $\tau$ and defining 
\begin{equation}
\frac{d \rho_S}{d t} = \lim\limits_{\tau \to 0} \frac{\rho_S' - \rho_S}{\tau}, 
\end{equation}
then  leads to Eq.~(\ref{M}).

Below we will also need the updated state of the ancillae, after they have interacted with the system. 
This can be  obtained by taking the partial trace of Eq.~(\ref{SM_unitary_expansion}) over $S$. 
Keeping only terms which are at most linear in $\tau$, we find
\begin{equation}\label{SM_rhoA_prime}{
\rho_A' = \rho_A^\text{th} + \sqrt{\tau} \bigg(\lambda \chi_A - i [G_{A}, \rho_A^\text{th}]\bigg) + \tau \bigg(- i \lambda  [G_{A}, \chi_A] + D_A(\rho_A^\text{th})\bigg).
}\end{equation}
where 
\begin{IEEEeqnarray}{rCl}
\label{SM_GA}
G_A &=& \tr_S (V_{SA} \rho_S), \\[0.2cm]
\label{SM_DA}
D_A(\rho_A^\text{th}) &=& -\frac{1}{2} \tr_S [V_{SA}, [V_{SA}, \rho_S \rho_A^\text{th}]]. 
\end{IEEEeqnarray}
Quite relevant to the discussion below, the term of order $\sqrt{\tau}$  does not vanish in Eq~(\ref{SM_rhoA_prime}).

\subsection*{Energy balance}

Using the general map~(\ref{SM_rhoSA_prime}) we can compute the changes in energy of the system and ancilla, defined as $\Delta H_S = \tr\big\{ H_S (\rho_{SA}'- \rho_{S} \rho_{A})\big\}$ and $\Delta H_A = \tr\big\{ H_A (\rho_{SA}'- \rho_{S} \rho_{A})\big\}$.
One then readily finds 
\begin{IEEEeqnarray}{rCl}
\label{SM_Delta_HS}
\Delta H_S &=& i \sqrt{\tau} \langle [V_{SA}, H_S] \rangle_{SA} - \frac{\tau}{2} \langle [V_{SA}, [V_{SA}, H_S]] \rangle_{SA},	\\[0.2cm]
\label{SM_Delta_HA}
\Delta H_A &=& i \sqrt{\tau} \langle [V_{SA}, H_A] \rangle_{SA} - \frac{\tau}{2} \langle [V_{SA}, [V_{SA}, H_A]] \rangle_{SA}.
\end{IEEEeqnarray}
where $\langle \ldots \rangle_{SA}$ means averages over $\rho_{S} \rho_A$.
In writing these formulas, we have not yet specified the state $\rho_A$ in order to emphasize 
the fact that the structure of these results is entirely independent of it. 
Due to strong energy conservation, $[V_{SA}, H_S + H_A] = 0$ it follows that 
\begin{IEEEeqnarray}{rCl}
\label{SM_energy_balace_Wc}
\langle [V_{SA}, H_S] \rangle_{SA} &=& - \langle [V_{SA}, H_A] \rangle_{SA},	\\[0.2cm]
\label{SM_energy_balace_Qinc}
\langle [V_{SA}, [V_{SA}, H_S]] \rangle_{SA} &=& -\langle [V_{SA}, [V_{SA}, H_A]] \rangle_{SA},
\end{IEEEeqnarray}
and hence 
\begin{equation}
\Delta H_S = - \Delta H_A. 
\end{equation}
Two conclusions may be drawn from this. 
The first is that, as  mentioned in the main text, the strong energy conservation condition~(4) implies that no work is performed; all change in energy in the system stems from a corresponding change in the ancillae. 
Second, Eqs.~(\ref{SM_energy_balace_Wc}) and (\ref{SM_energy_balace_Qinc}) allow us to pinpoint the origin of the coherent work $\mathcal{W}_C$ and the incoherent heat $\mathcal{Q}_\text{inc}$ appearing in Eq.~(\ref{energy_balance_MEq}) of the main text. 

To accomplish this, we simply need to express global averages over $\rho_S\rho_A$ in terms of local averages over either $\rho_S$ or $\rho_A$. 
For instance, referring to Eq.~(\ref{SM_Delta_HS}), the first term is precisely the coherent work since 
\[
\mathcal{W}_C = i \lambda \tau \langle [G, H_S]\rangle_S = i \sqrt{\tau} \langle [V_{SA}, H_S] \rangle_{SA}.
\]
The identity in Eq.~(\ref{SM_energy_balace_Wc}) therefore implies that this contribution will stems from a corresponding term on the side of the ancilla of the  form $\langle [V_{SA}, H_A] \rangle$. Whence,
\begin{equation}
\mathcal{W}_C = -i \sqrt{\tau} \langle [V_{SA}, H_A] \rangle_{SA} = -i \sqrt{\tau} \langle [G_A, H_A] \rangle_{A} , 
\end{equation}
where $G_A$ is given in Eq.~(\ref{SM_GA}).
In the last term, the average over $\rho_A^\text{th}$ does not contribute, so we finally get
\begin{equation}\label{SM_WC}
\mathcal{W}_C =-i \lambda \tau \langle [G_A, H_A] \rangle_{\chi_A}. 
\end{equation}
Similarly, the incoherent heat $\mathcal{Q}_\text{inc}$  is related to the second term in Eq.~(\ref{SM_Delta_HS}):
\begin{equation}\label{SM_Qinc_ancilla_side}
\mathcal{Q}_\text{inc} = \tau \tr\bigg\{ H_S D(\rho_S)\bigg\}  = - \frac{\tau}{2} \langle [V_{SA}, [V_{SA}, H_S]] \rangle = \frac{\tau}{2} \langle [V_{SA}, [V_{SA}, H_A]] \rangle = -\tau \tr\bigg\{ H_A D_A(\rho_A^\text{th})\bigg\},
\end{equation}
where $D_A(\rho_A^\text{th}) = -\frac{1}{2} \tr_S [V_{SA}, [V_{SA}, \rho_S \rho_A^\text{th}]]$.
The total change in energy of the system, which is the heat leaving the ancilla, can then be written solely in terms of ancilla-based quantities:
\begin{equation}\label{SM_heat_ancilla}{
\Delta \langle H_S \rangle := - Q_A = \mathcal{W}_C + \mathcal{Q}_\text{inc} = -i \lambda \tau \langle [G_A, H_A] \rangle_{\chi_A}-\tau \tr\bigg\{ H_A D_A(\rho_A^\text{th})\bigg\}.
}\end{equation}
These results therefore allow us to pinpoint which terms in the heat $\Delta H_A$ leaving the ancilla are converted to $\mathcal{W}_C$ and $\mathcal{Q}_\text{inc}$.

%
%
\section{S2. Perturbative expansion of entropic quantities}
%
%

The states of system and ancilla, before and after the interactions, will generally depend on $\tau$ in different ways.
To compute the entropy production, defined in Eq~(\ref{second_law}) of the main text,  one must compute several entropic quantities depending on these states.
Since we are interested in the limit $\tau \to 0$, these quantities can be computed using perturbation theory, which becomes \emph{exact} in the limit $\tau \to 0$. 
In this section we start by stating some general results on perturbative expansions of the von Neumann entropy, the relative entropy of coherence and the quantum Kullback-Leibler divergence (relative entropy). 
In Sec.~S3 we will then specialize these results to the relevant states appearing in Eq. (\ref{second_law}) and derive the results in Eqs. (\ref{mutual_info_result}) and (\ref{relative_entropy_result}).

\subsection{Von Neumann entropy}

Consider a general density matrix of the form 
\begin{equation}\label{SM_state_structure}
\rho = \rho_0 + \epsilon \sigma, 
\end{equation}
where $\epsilon$ is a small parameter and we assume $\tr \rho_0 = 1$ so $\tr \sigma = 0$. 
Let $\rho_0 = \sum_i p_i |i\rangle\langle i|$ denote the eigendecomposition of the unperturbed density matrix $\rho_0$. 
We now wish to compute the von Neumann entropy of $\rho$, which reads
\begin{equation}\label{SM_vonNeumann_def}
S(\rho) = - \tr(\rho \ln \rho) = - \sum\limits_i P_i \ln P_i, 
\end{equation}
where $P_i$ are the eigenvalues of the full density matrix $\rho$. 

Since $\rho$ is a Hermitian operator, standard perturbation theory applies \cite{Chen2010a}. 
Assuming that the $p_i$ are non-degenerate, we may then write, up to order $\epsilon^2$, 
\begin{equation}\label{SM_pert_eigenvalues}
P_i = p_i + \epsilon \sigma_{ii} + \epsilon^2 \sum\limits_{j\neq i} \frac{|\sigma_{ij}|^2}{p_i-p_j}.
\end{equation}
Plugging this in Eq.~(\ref{SM_vonNeumann_def}), expanding $P_i \ln P_i$ in $\epsilon$ up to second order and using the fact that $\tr \sigma = 0$, we find that
\begin{equation}\label{SM_vonNeumann_series}{
S(\rho) = S(\rho_0) - \epsilon \sum\limits_i \sigma_{ii} \ln p_i - \epsilon^2 \sum\limits_{i}\bigg\{ \frac{\sigma_{ii}^2}{2p_i} + \sum\limits_{j\neq i} \frac{|\sigma_{ij}|^2}{p_i-p_j} \ln p_i\bigg\}.
}\end{equation}
This is the series expansion for $S(\rho)$. The populations $\sigma_{ii}$ contribute both with order $\epsilon$ and $\epsilon^2$, whereas the coherences (off-diagonals) only start to contribute at order $\epsilon^2$.

\subsection{Relative entropy of coherence}

Due to this separation, the relative entropy of coherence [cf. the definition below Eq.~(15) of the main text] will be of order $\epsilon^2$: 
\begin{equation}\label{SM_rel_entropy_coherence}
\mathcal{C}(\rho)  = \epsilon^2 \sum\limits_{i,j\neq i} \frac{|\sigma_{ij}|^2}{p_i-p_j} \ln p_i.
\end{equation}
This expression can also be written more symmetrically, as 
\begin{equation}\label{SM_rel_entropy_coherence_2}
\mathcal{C}(\rho)  = \frac{\epsilon^2}{2} \sum\limits_{i,j\neq i} \frac{|\sigma_{ij}|^2}{p_i-p_j} \ln p_i/p_j.
\end{equation}
Thus, we see that the relative entropy of coherence weights each coherence $|\sigma_{ij}|$ by a factor of the form 
\[
\frac{\ln(x/y)}{x-y} \geq 1, \qquad x,y \in [0,1].
\]

\subsection{Quantum relative entropy}

Next we consider the relative entropy $S(\rho'||\rho)$ between two density matrices of the form 
\begin{equation}\label{structure_relative_entropy}
\rho = \rho_0 + \epsilon \sigma, \qquad
\rho' = \rho_0 + \epsilon \mu,
\end{equation}
where $\sigma$ and $\mu$ are arbitrary, but both depend on $\rho_0$ to order $\epsilon^0$. 
We have 
\begin{equation}
S(\rho'|| \rho) = - S(\rho') - \tr(\rho' \ln \rho). 
\end{equation}
The first term was already found in~(\ref{SM_vonNeumann_series}), with $\sigma$ replaced by $\mu$:
\begin{equation}\label{SM_Entropy_mu}
S(\rho') = S(\rho_0) - \epsilon \sum\limits_i \mu_{ii} \ln p_i - \epsilon^2 \sum\limits_{i}\bigg\{ \frac{\mu_{ii}^2}{2p_i} + \sum\limits_{j\neq i} \frac{|\mu_{ij}|^2}{p_i-p_j} \ln p_i\bigg\}.
\end{equation}

In order to compute the last term we will need not only the perturbation theory for the eigenvalues of $\rho$ [Eq.~(\ref{SM_pert_eigenvalues})], but also for its eigenvectors.
Defining $\rho = \sum\limits_i P_i |\tilde{i}\rangle\langle \tilde{i} |$ allows us to write
\begin{equation}\label{SM_KL_weird_term}
\tr(\rho' \ln\rho) = \sum\limits_i \langle \tilde{i} | \rho' | \tilde{i} \rangle \ln P_i.
\end{equation}
Thus, in addition to writing $\ln P_i$ as a power series, we will also have to expand $\langle \tilde{i} | \rho' | \tilde{i} \rangle$. 

Using standard perturbation theory, the eigenvectors of $\rho$ can be written as 
\begin{equation}\label{SM_pert_eigenvectors}
|\tilde{i}\rangle = |i\rangle + \epsilon |i_1\rangle + \epsilon^2 |i_2\rangle, 
\end{equation}
where 
\begin{IEEEeqnarray}{rCl}
\label{SM_pert_eigenvectors_i1}
|i_1\rangle &=& \sum\limits_{j\neq i} |j\rangle \frac{\sigma_{ij}}{p_i - p_j},		\\[0.2cm]
\label{SM_pert_eigenvectors_i2}
|i_2\rangle &=& 
- \frac{1}{2} |i\rangle \sum\limits_{j\neq i} \frac{|\sigma_{ij}|^2}{(p_i - p_j)^2}
- \sum\limits_{j\neq i} |j\rangle \frac{\sigma_{ii} \sigma_{ji}}{(p_i - p_j)^2}
+\sum\limits_{j\neq i, k \neq i} |k\rangle  \frac{\sigma_{kj} \sigma_{ji}}{(p_i-p_j)(p_i-p_k)} 
\end{IEEEeqnarray}
With this we find, after carrying out the computations, 
\begin{equation}
\langle \tilde{i} | \rho' | \tilde{i} \rangle = p_i + \epsilon \mu_{ii} + \epsilon^2 \sum\limits_{j\neq i} \frac{\mu_{ij} \sigma_{ji} + \sigma_{ij} \mu_{ji} - |\sigma_{ij}|^2}{p_i - p_j}.
\end{equation}

Plugging this result in Eq.~(\ref{SM_KL_weird_term}) and expanding all terms in $\epsilon$ then finally leads to 
\begin{equation}\label{SM_weird_series}
\tr(\rho' \ln\rho) = \sum\limits_i \Bigg\{p_i \ln p_i + \epsilon (\sigma_{ii} + \mu_{ii} \ln p_i) + \epsilon^2 \bigg[ \frac{\mu_{ii} \sigma_{ii}}{p_i}- \frac{\sigma_{ii}^2}{2 p_i} 
+ \sum\limits_{j\neq i}  \frac{\mu_{ij} \sigma_{ji} + \sigma_{ij} \mu_{ji} - |\sigma_{ij}|^2}{p_i - p_j} \ln p_i  \bigg] \Bigg\},
\end{equation}
which is correct up to order $\epsilon^2$.
Finally, combining this with Eq.~(\ref{SM_Entropy_mu}) leads to 
\begin{equation}\label{SM_KL_series}{
S(\rho' || \rho) = \frac{\epsilon^2}{2} \sum\limits_{i} \Bigg\{ \frac{(\mu_{ii} - \sigma_{ii})^2}{p_i} + \sum\limits_{j\neq i} \frac{|\mu_{ij} - \sigma_{ij}|^2}{p_i - p_j} \ln(p_i/p_j)\Bigg\}. 
}\end{equation}
We therefore see that while $S(\rho)$ and $S(\rho')$ contain contributions of order $\epsilon$, the first non-zero contribution to the relative entropy is of order $\epsilon^2$. 
Moreover, the result depends on both the populations and the coherences, and both with the same order $\epsilon^2$. 
This highlights some of the differences between $S(\rho'||\rho)$ and $S(\rho') - S(\rho)$.

%
%
\section{S3. Calculation of $\mathcal{I}(\rho_{SA}')$ and $S(\rho_{A}'||\rho_{A})$ [Eqs. (\ref{mutual_info_result}) and (\ref{relative_entropy_result}) of the main text]}
%
%

We are now in the position to derive Eqs. (\ref{mutual_info_result}) and (\ref{relative_entropy_result}) of the main text. 
To do so, one must simply apply the results of Sec. S2 with the appropriate choices of $\rho$, $\rho_0$, etc. 
Since system and environment always start uncorrelated and since the global dynamics is unitary,  the mutual information developed in the map~(\ref{SM_rhoSA_prime}) can be written as 
\begin{equation}\label{SM_MI_def}
\mathcal{I}(\rho_{SA}') = S(\rho_S') + S(\rho_A') - S(\rho_{SA}') := \Delta S_S + \Delta S_A.
\end{equation}
Our task is to compute $\Delta S_A = S(\rho_A') - S(\rho_A)$. 
In addition, we will also need $S(\rho_A' ||\rho_A)$. 
We compute each term separately.

\subsection{Calculation of $S(\rho_A)$}

The initial state of the ancilla is given in Eq.~(\ref{ancilla_state}) of the main text, $\rho_A = \rho_A^\text{th}+ \sqrt{\lambda^2 \tau} \chi_A$, where $\chi_A$ has no diagonal elements. 
This falls under the structure of Eq.~(\ref{SM_state_structure}), provided we identify 
\[
\epsilon \sigma = \sqrt{\tau} \lambda \chi_A.
\]
A direct application of Eq.~(\ref{SM_vonNeumann_series}) then yields
\begin{equation}\label{SM_S_A_step1}
S(\rho_A) = S(\rho_A^\text{th}) - \lambda^2 \tau \sum\limits_{i,j\neq i}  \frac{|(\chi_A)_{ij}|^2}{p_i^\text{th} - p_j^\text{th}} \ln p_i^\text{th}, 
\end{equation}
where $p_i^\text{th}$ are the eigenvalues of $\rho_A^\text{th}$ and the basis $|i\rangle$ refers to the energy basis of $H_A$.
Since the perturbed part of $\rho_A$ has no diagonal elements, the second term in Eq.~(\ref{SM_S_A_step1}) is nothing but the relative entropy of coherence of the state $\rho_A$,
\begin{equation}\label{SM_C_rhoA}
C(\rho_A) = \lambda^2 \tau \sum\limits_{i,j\neq i}  \frac{|(\chi_A)_{ij}|^2}{p_i^\text{th} - p_j^\text{th}} \ln p_i^\text{th}.
\end{equation}
Thus, we may simply write 
\begin{equation}\label{SM_S_rhoA}{
S(\rho_A) = S(\rho_A^\text{th}) - C(\rho_A).
}\end{equation}

\subsection{Calculation of $S(\rho_A')$}

The state of the ancilla after the map is given by Eq.~(\ref{SM_rhoA_prime}).  This once again has the structure Eq.~(\ref{SM_state_structure}), 
but now one must identify 
\[
\epsilon \sigma = \sqrt{\tau}\bigg\{\lambda  \chi_A - i [G_{A}, \rho_A^\text{th}] \bigg\} + \tau \bigg\{ - i \lambda  [G_{A}, \chi_A] + D_A(\rho_A^\text{th})\bigg\}.
\]
The terms proportional to $\sqrt{\tau}$ now form the off-diagonal part of $\sigma$ and those proportional to $\tau$ are all diagonal. 
Applying again Eq.~(\ref{SM_vonNeumann_series}) yields
\[
S(\rho_A') = S(\rho_A^\text{th}) - \tau \sum\limits_i \bigg( -i \lambda [G_A, \chi_A] + D_A(\rho_A^\text{th})\bigg)_{ii} \ln p_i^\text{th} - \tau \sum\limits_{i,j\neq i} \frac{\bigg|\bigg(\lambda \chi_A - i [G_A, \rho_A^\text{th}]\bigg)_{ij}\bigg|^2}{p_i^\text{th}-p_j^\text{th}} \ln p_i^\text{th}.
\]
Once again, comparing with Eqs.~(\ref{SM_vonNeumann_series}) and (\ref{SM_rel_entropy_coherence}), the relative entropy of coherence of $\rho_A'$ corresponds to the last term only,
\begin{equation}\label{SM_C_rhoAprime}
C(\rho_A') = \tau \sum\limits_{i,j\neq i} \frac{\bigg|\bigg(\lambda \chi_A - i [G_A, \rho_A^\text{th}]\bigg)_{ij}\bigg|^2}{p_i^\text{th}-p_j^\text{th}} \ln p_i^\text{th}.
\end{equation}
That is, we may write 
\[
S(\rho_A') = S(\rho_A^\text{th}) - \tau \sum\limits_i \bigg( -i \lambda [G_A, \chi_A] + D_A(\rho_A^\text{th})\bigg)_{ii} \ln p_i^\text{th} - C(\rho_A'). 
\]
The  term proportional to $\tau$, on the other hand, can be written as 
\[
 - \tau \sum\limits_i \bigg( -i \lambda [G_A, \chi_A] + D_A(\rho_A^\text{th})\bigg)_{ii} \ln p_i^\text{th}  = \beta\tau \tr_A \bigg\{ \bigg( -i \lambda [G_A, \chi_A] + D_A(\rho_A^\text{th})\bigg) H_A\bigg\},\
\]
where we also use the fact that $\ln \rho_A^\text{th} = - \beta H_A - \ln Z_A$ [c.f. Eq.~(\ref{ancilla_state}) of the main text]. 
The terms proportional to $\ln Z_A$ vanish since the operators in the above expression are all traceless. 
This term is therefore nothing but the total change in energy of the ancilla in Eq.~(\ref{SM_heat_ancilla}). 
But this, in turn, is minus the change in energy in the system.  
Whence, we conclude that 
\begin{equation}\label{SM_S_rhoA_prime}{
S(\rho_A') = S(\rho_A^\text{th}) - \beta \Delta \langle H_S \rangle - C(\rho_A').
}\end{equation}

\subsection{Calculation of $\mathcal{I}(\rho_{SA}')$}

Inserting Eqs.~(\ref{SM_S_rhoA}) and (\ref{SM_S_rhoA_prime}) into Eq.~(\ref{SM_MI_def}) leads to 
\begin{equation}{
\mathcal{I}(\rho_{SA}') = \Delta S_S - \beta \Delta \langle H_S \rangle - \Delta C_A,
}\end{equation}
which is Eq.~(\ref{mutual_info_result}) of the main text, provided we recognize $\Delta S_S - \beta \Delta \langle H_S \rangle = - \beta \Delta F$, as the change in free energy of the system. 

\subsection{Calculation of $S(\rho_A'||\rho_A)$}

Finally,  we turn to the relative entropy $S(\rho_A'||\rho_A)$, expressed as the series in Eq.~(\ref{SM_KL_series}).
The operators $\mu$ and $\sigma$, defined in Eq.~(\ref{structure_relative_entropy}), should now be recognized with 
\begin{IEEEeqnarray*}{rCl}
\epsilon \sigma &=& \sqrt{\tau} \lambda \chi_A, \\[0.2cm]
\epsilon \mu &=& \sqrt{\tau}\bigg\{\lambda  \chi_A - i [G_{A}, \rho_A^\text{th}] \bigg\} + \tau \bigg\{ - i \lambda  [G_{A}, \chi_A] + D_A(\rho_A^\text{th})\bigg\}.
\end{IEEEeqnarray*}
The first term in Eq.~(\ref{SM_KL_series}) will depend only on the diagonal part of $\mu$ (the diagonal part of $\sigma$ is zero). 
But this term is already of order $\tau$, so this will ultimately lead to a contribution of order $\tau^2$. 

The only non-negligible term is thus the one related to the coherences. 
It is convenient to express $|\mu_{ij} - \sigma_{ij}|^2$ as 
\[
|\mu_{ij} - \sigma_{ij}|^2 = |\mu_{ij}|^2 - |\sigma_{ij}|^2 + \bigg( 2 |\sigma_{ij}|^2 - \mu_{ij} \sigma_{ji} - \sigma_{ij} \mu_{ji}\bigg). 
\]
The reason why this is useful is because then the first two terms can be recognized as the  difference between the relative entropies of coherence of $\rho_A'$ and $\rho_A$ respectively [Eqs.~(\ref{SM_C_rhoAprime}) and (\ref{SM_C_rhoA})]. 
On the other hand, the remaining term in parenthesis may be written as 
\[
2 |\sigma_{ij}|^2 - \mu_{ij} \sigma_{ji} - \sigma_{ij} \mu_{ji} = i \lambda \bigg[ (\chi_A)_{ij} (G_A)_{ji} - (G_A)_{ij} (\chi_A)_{ji} \bigg] (p_i^\text{th} - p_j^\text{th}).
\]
Substituting these results in Eq.~(\ref{SM_KL_series}) and expressing the remaining summations in terms of a trace, then yields
\[
S(\rho_A' || \rho_A) = \Delta C_A - i \lambda \beta \tau \langle [G_A,H_A] \rangle_{\chi_A}.
\]
Comparing this  with Eq.~(\ref{SM_WC}), we finally arrive at Eq.~(\ref{relative_entropy_result}) of the main text; viz.,
\begin{equation}{
S(\rho_A' || \rho_A) = \Delta C_A + \beta \mathcal{W}_C.
}\end{equation}

%
%
\section{S4. Ergotropy}
%
%

Using the results of Sec.~S.2 it is straightforward to compute the ergotropy of the ancilla state $\rho_A$ (see main text for discussions). 
To conform with the notation of Sec.~S.2., let us study instead the state~(\ref{SM_state_structure}), but assuming that $\rho_0$ is a thermal state $\rho_0 = e^{-\beta H}/Z$, with population $p_i = e^{-\beta E_i}/Z$ and eigenbasis $|i\rangle$. 
Matching this to the notation of the ancilla state $\rho_A$ [Eq.~(5) of the main text] will be straightforward once the final result has been derived. 

The ergotropy is defined as (Ref.~\cite{Allahverdyan2004} of the main text)
\begin{equation}\label{SM_ergotropy_def}
\mathcal{W} = \sum\limits_{ij} P_i E_j \bigg[ |\langle j | \tilde{i} \rangle|^2 - \delta_{ij}\bigg].
\end{equation}
Here $P_i$ and $|\tilde{i}\rangle$ are the eigenvalues and eigenvectors of $\rho$, which are given by Eqs.~(\ref{SM_pert_eigenvalues}) and~(\ref{SM_pert_eigenvectors}) respectively. 
Using the fact that $\langle i | i_1 \rangle = 0$ [c.f.~Eq.~(\ref{SM_pert_eigenvectors_i1})], it follows that 
\[
|\langle j | \tilde{i} \rangle|^2  = \delta_{ij} + \epsilon^2 \bigg[ |\langle j | i_1 \rangle|^2 + 2\delta_{ij} \text{Re} (\langle i | i_2 \rangle) \bigg],
\]
which is thus already of order $\epsilon^2$. 
We now  plug this in Eq.~(\ref{SM_ergotropy_def}).
Restricting the calculation only up to $\epsilon^2$, as was done in the previous sections, we see that we only need to take the zeroth order term of $P_i$ [Eq.~(\ref{SM_pert_eigenvalues})], which is precisely the equilibrium probabilities $p_i$. 
Whence, Eq.~(\ref{SM_ergotropy_def}) becomes
\[
\mathcal{W} = \epsilon^2 \sum\limits_{ij} p_i E_j \bigg[|\langle j | i_1 \rangle|^2 + 2\delta_{ij} \text{Re} (\langle i | i_2 \rangle) \bigg].
\]
From~(\ref{SM_pert_eigenvectors_i1}) and (\ref{SM_pert_eigenvectors_i2}) we get 
\begin{IEEEeqnarray*}{rCl}
|\langle j | i_1 \rangle|^2 &=& (1-\delta_{ij}) \frac{|\sigma_{ij}|^2}{(p_i - p_j)^2},		\\[0.2cm]
2 \text{Re}(\langle i | i_2 \rangle) &=& - \sum\limits_{j\neq i} \frac{|\sigma_{ij}|^2}{(p_i - p_j)^2}.
\end{IEEEeqnarray*}
This allows us to write 
\[
\mathcal{W} = \epsilon^2 \sum\limits_{i, j\neq i} p_i (E_j-E_i) \frac{|\sigma_{ij}|^2}{(p_i - p_j)^2}.
\]
Finally, we write $E_j - E_i = T \ln p_i/p_j$ and also symmetrize the formula, exactly as was done in going from Eq.~(\ref{SM_rel_entropy_coherence}) to (\ref{SM_rel_entropy_coherence_2}). 
We then finally get 
\begin{equation}\label{ergotropy_final}
\mathcal{W} = \frac{T \epsilon^2}{2} \sum\limits_{i,j\neq i} \frac{|\sigma_{ij}|^2}{(p_i - p_j)}\ln p_i/p_j = T C(\rho),
\end{equation}
where the connection with the relative entropy of coherence  $C(\rho)$ can be directly verified by comparing with  Eq.~(\ref{SM_rel_entropy_coherence_2}). 
Translating this now to the weakly coherent state of the ancilla [Eq.~(5) of the main text], one concludes that the ergotropy contained in $\rho_A$ is simply $\mathcal{W} = T C(\rho_A)$.

\end{document}